\documentclass[twocolumn,prX,nofootinbib]{revtex4-1}
\pdfoutput=1
\usepackage{amsmath,amssymb,amsfonts,eucal,mathrsfs,graphicx,tensor}
\usepackage[usenames,dvipsnames]{color}
\usepackage{hyperref}
\DeclareMathAlphabet{\mathpzc}{OT1}{pzc}{m}{it}

\newcommand{\warnb}[1]{{\color{blue}#1}}
\definecolor{brown}{rgb}{0.65,0.16,0.16}\newcommand{\warnbr}[1]{{\color{brown}#1}}
\newcommand{\warng}[1]{{\color{OliveGreen}#1}}

\newcommand{\beqn}[1]{\begin{equation*}#1\end{equation*}}
\newcommand{\beql}[1]{\begin{equation}\label{#1}}
\newcommand{\eeq}{\end{equation}}
\newcommand{\eq}[1]{\eqref{#1}}

\newcommand{\iim}{\Leftrightarrow}

\newcommand{\lf}{\left}
\newcommand{\rf}{\right}

\newcommand{\ft}{\protect\footnote}

\newcommand{\rt}{\sqrt}
\newcommand{\fr}{\frac}
\newcommand{\tn}{\tensor}

\newcommand{\ord}{{\boldsymbol{\CMcal{O}}}}

\newcommand{\de}{\delta}

\newcommand{\tH}{\theta}
\newcommand{\ka}{\kappa}
\newcommand{\la}{\lambda}
\newcommand{\La}{\Lambda}
\newcommand{\Sg}{\Sigma}

\newcommand{\Om}{\Omega}

\newcommand{\wg}{{\,\wedge\;}}

\newcommand{\ed}{\mathrm{d}}
\newcommand{\dl}{\partial}
\newcommand{\Dl}{\nabla}
\newcommand{\vecDl}{\vec{\nabla}}

\newcommand{\dr}{\mathrm{d}r}
\newcommand{\ds}{\mathrm{d}s}
\newcommand{\du}{\mathrm{d}u}
\newcommand{\dU}{\mathrm{d}U}
\newcommand{\dv}{\mathrm{d}v}
\newcommand{\dtH}{\mathrm{d}\theta}
\newcommand{\dphi}{\mathrm{d}\phi}
\newcommand{\dOm}{\mathrm{d}\Omega}

\newcommand{\intdx}[1]{\int\mathrm{d}^{#1}x\,}

\newcommand{\met}{\mathsf{g}}
\newcommand{\pmet}{\mathsf{p}}
\newcommand{\hatmet}{\hat{\mathsf{g}}}

\newcommand{\veii}{\varepsilon^{\text{\sc ii}}}
\newcommand{\hatK}{\hat{K}}
\newcommand{\hatk}{\hat{k}}
\newcommand{\Rie}{\mathpzc{R}\,}
\newcommand{\Ric}{\mathpzc{R}\,}
\newcommand{\EinG}{\mathpzc{G}}

\newcommand{\Gae}{G_{\ae}}
\newcommand{\GN}{G_\text{\sc n}}

\newcommand{\ac}{\CMcal{S}}
\newcommand{\CC}{{\text{\sc cc}}}
\newcommand{\emT}{\mathpzc{T}}
\newcommand{\aeT}{\emT^{\ae}}
\newcommand{\aeTH}{\emT^{(\text{\sc h})}}
\newcommand{\aebT}{\mathbf{T}^{\ae}}
\newcommand{\aePi}{\vec{\Pi}^{\ae}}
\newcommand{\PiH}{\vec{\Pi}}
\newcommand{\vecAE}{\vec{\AE}}

\newcommand{\uhor}{\text{\sc uh}}
\newcommand{\ruh}{r_{\uhor}}

\newcommand{\eff}{\mathrm{eff}}
\newcommand{\elef}{\ell_{\eff}}

\newcommand{\lag}{\mathscr{L}}
\newcommand{\aelagcon}{\lag_{\ae}^{(\textsc{con})}}

\newcommand{\qsmarr}{q_{\textsc{smarr}}}
\newcommand{\quh}{q_\uhor}
\newcommand{\kauh}{\kappa_\uhor}
\newcommand{\Xuh}{|\chi|_{\uhor}}

\newcommand{\Sph}{\mathscr{B}}

\begin{document}
\title{Universal horizons in maximally symmetric spaces}
\author{Jishnu Bhattacharyya}\email{jishnu.bhattacharyya@nottingham.ac.uk}
\affiliation{School of Mathematical Sciences, University of Nottingham, University Park, Nottingham, NG7 2RD, UK}
\author{David Mattingly}\email{dyo7@unh.edu}
\affiliation{Department of Physics, University of New Hampshire, Durham, NH 03824, USA}

\begin{abstract}
Universal horizons in Ho\v{r}ava-Lifshitz gravity and Einstein-{\ae}ther theory are the equivalent of causal horizons in general relativity and appear to have many of the same properties, including a first law of horizon thermodynamics and thermal radiation. Since universal horizons are infrared solutions of a putative power counting renormalizable quantum gravitational theory, fully understanding their thermodynamics will shed light on the interplay between black hole thermodynamics and quantum gravity. In this paper, we provide a complete classification, including asymptotic charges, of all four dimensional static and spherically symmetric universal horizon solutions with maximally symmetric asymptotics -- the equivalents of the Schwarzschild, Schwarzschild de Sitter or Schwarzschild anti-de Sitter spacetimes. Additionally we derive the associated first laws for the universal horizon solutions. Finally we prove that independent of asymptotic boundary conditions, any spherically symmetric solution in Horava-Lifshitz gravity with a universal horizon is also a solution of Einstein-{\ae}ther theory, thereby broadening and complementing the known equivalence region of the solution spaces.
\end{abstract}

\maketitle
\section{Introduction}
Black holes provide one of the few physical systems in which the direct application of two well tested physical theories, namely general relativity and quantum field theory, yields a robust, quantitative theoretical prediction about high energy quantum gravity: the number of quantum gravitational states of a macroscopic black hole is calculable and given by the Bekenstein-Hawking entropy (at lowest order). If we only knew of one solution for black holes, for example the four dimensional Schwarzschild solution introduced first in undergraduate relativity classes, then this prediction, while useful, would be a much less effective touchstone for quantum gravity, as it would provide only one number for one particular solution. However, there is a zoo of black hole solutions, including solutions with angular momentum, gauge field charges, non-spherical topologies, different dimensions, etc. In addition, corrections to the entropy can be computed both in low energy and quantum gravity theories. The combination of the black hole zoo and sophisticated technologies to compute the appropriate entropies has led to a vast literature on verifying quantum gravity theories by checking their agreement with the entropy of the known black hole solutions within the theory. The most success has come in string theory, although other quantum gravity theories have also achieved definite predictions.

The universality of various approaches to quantum gravity yielding the same black hole entropy led to the realization, originally due to Strominger~\cite{Strominger:SBH} and Carlip~\cite{Carlip:SBH}, that merely being near a Killing horizon (in a suitable sense) forces any putative quantum gravity states to admit a conformal field theory description. The conformal symmetry is enough to dictate the ultra-violet density of states and hence the entropy. As a result, black hole entropy (to leading order) is universal, in that if a black hole solution exists in the macroscopic limit of a quantum gravity theory, the entropy will come out appropriately. A related universality argument, although about holography rather than black hole entropy specifically, has been presented by Marolf~\cite{Marolf:gen-holo}. Here, one notes that in a generally covariant theory (with the key requirement being no a priori background structures) the on-shell Hamiltonian consists solely of boundary terms. As long as this property still holds in the quantum theory, then using properties of entanglement in quantum field theories one can construct a holographic dual field theory to a given gravitational bulk theory.

These two universality arguments highlight the important roles general covariance and the light cone structure of relativity play in black hole entropy calculations and corresponding holographic arguments. An interesting question now presents itself: given a causal horizon, are general covariance and light cone causal structures merely sufficient to have a notion of black hole entropy and holography, or are they actually necessary? To answer this question one needs to consider gravitational (or quantum gravitational) theories where there are solutions with a stationary, trapped region but the causal boundary is not a Killing horizon, or theories that are not generally covariant. At first sight, considering such theories seems perverse as one should not get rid of cherished principles lightly. However a recent proposal for quantum gravity~\cite{Horava:HL-intro}, Ho\v{r}ava-Lifshitz gravity, breaks the light cone structure and has seen a tremendous amount of interest (for a recent review, see~\cite{HL:review}). In addition, there is an extended version of Ho\v{r}ava-Lifshitz gravity~\cite{BPS:extHL} that, in the infrared (IR) limit (equivalently, in the two-derivative truncated limit), admits a re-description~\cite{Germani:2009yt, BPS:HL-xm-inconsistency,J:hso-ae=hor} in terms of the well-known Einstein-{\ae}ther theory~\cite{JM:ae:intro}, which has very different causal structures than general relativity.

In both Einstein-{\ae}ther and Ho\v{r}ava-Lifshitz theories, black hole solutions with causal boundaries, or `universal horizons'~\cite{BS:UH-instability, BJS:aebh}, distinct from Killing horizons exist \cite{EJ:aebh, BJS:aebh, BBM:mechuhor, JB:thesis, BS:aebh:R1, BS:aebh:R2, BS:aebh:RR, Janiszewski:2014iaa, Janiszewski:2014ewa, Sotiriou:2014gna}. Known solutions possess a first law of black hole mechanics~\cite{BBM:mechuhor, JB:thesis, Mohd:2013zca} and certain simple, analytically tractable solutions are known to radiate thermally from the universal horizon~\cite{BBM:thermo, JB:thesis, Cropp:2013sea}. However, neither a complete thermodynamical analysis nor corresponding quantum gravitational entropy calculation has been carried out, and indeed numerous open questions remain. It is therefore still unclear whether the local light cone structure is necessary for horizon thermodynamics.

In this paper we contribute to this discussion by classifying the spherically symmetric set of black hole solutions with maximally symmetric asymptotics for both extended Ho\v{r}ava-Lifshitz theory in the IR limit and the closely related Einstein-{\ae}ther theory. In this way, we increase the number of members of the black hole zoo for these theories, which allows for more quantitative and explicit checks of entropy (and other thermodynamic issues, e.g., black hole radiation, status of the second law, and so on) just as the black hole zoo does for general relativity and related field theories. We also revisit the relationship between the solution spaces of Ho\v{r}ava-Lifshitz versus Einstein-{\ae}ther theory, expanding on the comments of~\cite{J:hso-ae=hor} that showed that hypersurface orthogonal solutions in Einstein-{\ae}ther are also solutions of Ho\v{r}ava-Lifshitz theory.

The paper is organized as follows. In section~\ref{background}, we introduce both theories and present the relevant equations of motion. In section~\ref{ss:equiv} we prove that all static and spherically symmetric {\it black hole solutions} of Ho\v{r}ava-Lifshitz theory are also solutions of Einstein-{\ae}ther theory. This result therefore complements the more general claim of~\cite{J:hso-ae=hor} when confined within the subset of static and spherically symmetric solutions. The proof presented here does not rely on the asymptotics of the solutions, and is also independent of the presence of additional matter fields, as long as the solutions are static, spherically symmetric, and contain a black hole. It is therefore somewhat broader in applicability than similar proofs appearing in previous work~\cite{BS:UH-instability}, which require asymptotic behavior and are only applicable to asymptotically flat spacetimes where the {\ae}ther is aligned with the Killing vector at infinity. In particular, the proof in this paper can be trivially generalized to a broader class of solutions than considered in this paper, including solutions with flat transverse spaces, e.g., those that are commonly studied in the context of asymptotically Lifshitz solutions, or for solutions in $D \neq 4$, $D$ being the spacetime dimensionality.

In section~\ref{sss:aebh}, we perform a classification of static and spherically symmetric black hole solutions with maximally symmetric asymptotics. As a byproduct of our analysis, a number of static and spherically symmetric solutions, expressible in simple and closed analytic form, can be found when parameters in the Lagrangian are tuned to special and physically interesting values (these solutions have already appeared in~\cite{JB:thesis}). Our central result of this section can be summarized as follows: static, spherically symmetric and asymptotically maximally symmetric black hole solutions of Ho\v{r}ava-Lifshitz and Einstein-{\ae}ther theories fall in three categories:
	\begin{enumerate}
	\item the exact solutions appearing in~\cite{JB:thesis} (these solutions can have all types of maximally symmetric asymptotics),
	\item solutions with flat asymptotics numerically found previously in~\cite{EJ:aebh,BJS:aebh},
	\item solutions with de Sitter asymptotics, a preliminary study of which is presented in this work.
	\end{enumerate}
As an interesting corollary of our results, there are no anti-de Sitter black hole solutions besides those which can be expressed in closed exact form~\cite{JB:thesis}.

Finally in section~\ref{Smarr:1L}, we present the corresponding Smarr formula and first law of mechanics for the above solutions. The derivation of these formul{\ae} are presented in great details in~\cite{JB:thesis}, and our focus in the current presentation will be to revisit our introductory remarks on holography in these theories in light of these results.
\section{Background}\label{background}
We begin with a quick review of Einstein-{\ae}ther and Ho\v{r}ava-Lifshitz theories, especially focusing on the (dis)similarities between their equations of motion and the relationships between the corresponding solution spaces.
\subsection{Einstein-{\ae}ther theory}
Einstein-{\ae}ther theory~\cite{JM:ae:intro} is a generally covariant theory of a pair of dynamical fields: the metric $\met_{a b}$ and an `{\ae}ther' vector field $u^a$. The dynamics of this theory can be derived from an action which (modulo surface/boundary terms) is the standard Einstein-Hilbert action with a cosmological constant term plus a two-derivative action for the {\ae}ther
	\beql{ae:action}
	\ac = \fr{1}{16\pi\Gae}\intdx{4}\rt{-\met}\lf[-\fr{6c_{\CC}}{\ell^2} + \Rie + \lag_{\ae} + \aelagcon\rf]~.
	\eeq
Here $\ell$ is the length scale associated with the cosmological constant (which has been normalized canonically above), $c_{\CC} = -1$, $0$ and $1$ for negative, zero and positive cosmological constant, respectively, and
	\beqn{
	\aelagcon = \la_{\ae}(u^2 + 1)~,
	}
is a constraint term for the {\ae}ther with a Lagrange multiplier $\la_{\ae}$ that enforces the unit-norm constraint on the {\ae}ther vector (we use the mostly plus signature for the metric)
	\beql{ae:norm}
	u^2 = -1~.
	\eeq
General covariance is maintained by enforcing the unit constraint on $u^a$ as a dynamical constraint. Due to the tensor vacuum expectation value for the {\ae}ther field, every solution of the theory breaks local Lorentz invariance -- there is always a preferred observer defined by the {\ae}ther field.

The kinetic terms for the {\ae}ther are contained in the kinetic part of the Lagrangian $\lag_{\ae}$, which is given by
	\beqn{
	\lag_{\ae} = -\tn{Z}{^{a b}_{c d}}(\Dl_a u^c)(\Dl_b u^d)~,
	}
where the tensor
	\beqn{
	\tn{Z}{^{a b}_{c d}} = c_1\met^{a b}\met_{c d} + c_2\tn{\de}{^a_c}\tn{\de}{^b_d} + c_3\tn{\de}{^a_d}\tn{\de}{^b_c} - c_4u^a u^b\met_{c d}~,
	}
ensures all possible two derivative terms for the {\ae}ther\ft{Other seemingly possible terms, e.g., $\Ric_{a b} u^a u^b$, can be shown to be equivalent to the terms already presented above, up to total derivatives. Obviouly, such terms do not affect the local equations of motion.}. The four constants $c_1$, $c_2$, $c_3$ and $c_4$ play the role of coupling constants of the theory. In fact, the following linear combinations of these couplings will directly appear in our analysis
	\beqn{
	c_{13} = (c_1 + c_3)~, \quad \bar{c}_{13} = (c_1 - c_3)~, \quad c_{14} = (c_1 + c_4)~.
	}
In addition, the following two combinations will also turn out to be useful
	\beqn{
	c_{123} = (c_1 + c_2 + c_3)~, \quad c_l = \fr{2 + c_{13} + 3c_2}{2}~.
	}
We mention in passing that only $c_l$, among all the combinations defined above, gets a dimension dependent modification for $D \neq 4$. The constant $\Gae$ that normalizes the action is a dimensionful normalization constant required to make the action dimensionless. For asymptotically flat solutions, $\Gae$ may be related to $\GN$, i.e., Newton's gravitational constant, via
	\beqn{
	\Gae = \lf(1 - \fr{c_{14}}{2}\rf)\GN~,
	}
by considering the weak field/slow-motion limit of solutions of Einstein-{\ae}ther theory~\cite{Carroll:2004ai}.

In principle, one may find solutions of the equations of motion irrespective of the values of the $c_i$ couplings. However, such solutions will only be physically reasonable if they satisfy physicality conditions like positive energy, absence of naked singularities, etc. This is possible if the couplings are restricted to be within certain values. The limiting values correspond neatly with existing theoretical and observational bounds; see e.g.~\cite{J:ae:constraints, J:ae:status} for a comprehensive review. Specifically when analyzing physically reasonable solutions in this work, we will assume the following constraints to hold on the $c_i$ couplings
	\beql{c_i:constraints}
	c_{13} < 1~, \qquad 0 \leqq c_{14} < 2~, \qquad c_{123} \geqq 0~.
	\eeq
The constraints \eq{c_i:constraints} essentially come from demanding perturbative stability of the theory around flat space. To elaborate, if one considers linearized {\ae}ther-metric perturbations around flat spacetime in Einstein-{\ae}ther, one finds propagating spin-0, spin-1 and spin-2 modes, with respective propagation speeds $s_0$, $s_1$ and $s_2$ measured in the {\ae}ther's rest frame given by~\cite{JM:ae-waves}
	\beql{speeds}
	\begin{split}
	& s_0^2 = \fr{c_{123}}{c_{14}(1 - c_{13})c_l}\lf(1 - \fr{c_{14}}{2}\rf)~, \\
	& s_1^2 = \fr{c_{13} + \bar{c}_{13} - c_{13}\bar{c}_{13}}{2(1 - c_{13})c_{14}}~, \qquad s_2^2 = \fr{1}{1 - c_{13}}~.
	\end{split}
	\eeq
The requirement of perturbative stability then translates to demanding reality of the various speeds above. In particular, reality of the spin-2 and the spin-0 mode speeds gives us all the constraints presented in~\eq{c_i:constraints}. The bounds on $c_{13}$ and $c_{123}$ further imply
	\beqn{
	c_l > 0~,
	}
which will be very important in the context of non-trivial asymptotic behavior of the {\ae}ther in some of the physically relevant solutions we present later. The only additional constraint that is not evident from the perturbative analysis is that $c_{14} = 2$ is also forbidden due to the requirement that solutions with a non-trivial {\ae}ther/matter profile have positive definite total mass.

We do not require any restriction on $\bar{c}_{13}$ in this work, because, we are eventually going to specialize to a sector of the Einstein-{\ae}ther theory solution space where the {\ae}ther is hypersurface orthogonal\ft{In this case, the spin-1 mode of {\ae}ther perturbations, related to the twist of the {\ae}ther field, cannot survive. The dependence of the spin-1 mode speed~\eq{speeds} on $\bar{c}_{13}$, the coefficient of the `twist-squared' term in the action~\eq{ae:action}, is a direct indication of this. When the {\ae}ther is hypersurface orthogonal, and hence twist free, all reference to $\bar{c}_{13}$ drops out. We elaborate on this a little more below; for a more detailed discussion, see~\cite{J:hso-ae=hor, J:undo-w}.}.

There are other observational limits on the couplings, e.g., coming from the requirement that propagating high energy cosmic rays do not lose energy due to vacuum \v{C}erenkov radiation of gravitons~\cite{Elliott:2005va}. We will explicitly not impose such constraints here as we are interested in the behavior of the scalar mode, and hence allowing the scalar mode to have any speed consistent with~\eq{c_i:constraints} is theoretically useful.

Extremizing the action~\eq{ae:action} under variations of the metric and the {\ae}ther, one is led to the following equations of motion of Einstein-{\ae}ther theory
	\beql{ae:EOM}
	\EinG_{a b} = -\fr{3c_{\CC}}{\ell^2}\met_{a b} + \aeT_{a b}~, \qquad \AE_a = 0~,
	\eeq
where, $\EinG_{a b}$ is the Einstein tensor, while the {\ae}ther stress tensor $\aeT_{a b}$ and $\AE_a$ (the functional derivative of the {\ae}ther action with respect to $u^a$) are defined (up to boundary terms) as follows
	\beqn{
	\de\lf[\rt{-\met}(\lag_{\ae} + \aelagcon)\rf] = \rt{-\met}\lf(-\aeT_{a b} \de\met^{a b} + 2\AE_a \de u^a\rf)~.
	}
The explicit expressions for $\aeT_{a b}$ and $\AE_a$ for the most general case, in terms of derivatives of the metric and the {\ae}ther, can be found in the standard literature, e.g.,~\cite{JM:ae:intro}. We will reduce the equations of motion to the spherically symmetric case and write out the explicit forms further on. The full dynamical equations~\eq{ae:EOM} along with the unit-norm constraint~\eq{ae:norm} form the complete set of equations of motion of Einstein-{\ae}ther theory.
\subsection{Ho\v{r}ava-Lifshitz theory in the IR limit}\label{background:HL}
The {\ae}ther action in the previous section is closely related to the non-projectable version of Ho\v{r}ava-Lifshitz gravity developed by Blas, Pujolas, and Sibiryakov (BPS)~\cite{BPS:extHL}. In the `covariant formulation' of the BPS extension, the foliation is dynamical and generated by a scalar time function $U$. In order that $U$ serves no purpose other than to label the leaves of the foliation, the action can only depend on $U$ via the co-vector
	\beql{ae:HSO}
	u_a = -N\Dl_a U~, \quad N^{-2} = -\met^{a b}(\Dl_a U)(\Dl_b U)~,
	\eeq
where the second relation takes care of the unit norm constraint~\eq{ae:norm} on the {\ae}ther. This ensures that the {\ae}ther (co)vector (and therefore any quantity that is defined in terms of it, including the action) is invariant under the following reparameterizations of $U$ that preserve the foliation
	\beql{U:reparams}
	U \mapsto \tilde{U}(U)~, \qquad N \mapsto \tilde{N} = (\ed\tilde{U}/\dU)^{-1}N~.
	\eeq
If one were to construct a coordinate system adapted to the foliations defind by the {\ae}ther, which could be referred to as the preferred frame, then $U$ will play the role of time in that frame, while $N$ will be the preferred frame lapse function.

We can rewrite the Ho\v{r}ava-Lifshitz theory action in terms of $u_a$ explicitly, making clear the link between Einstein-{\ae}ther and Ho\v{r}ava-Lifshitz gravity. In fact, the action for the non-projectable version of Ho\v{r}ava-Lifshitz gravity is identical with~\eq{ae:action} without the Lagrange multiplier term\ft{The definition~\eq{ae:HSO} of the {\ae}ther in terms of the scalar function $U$ automatically ensures that $u_a$ is unit. In fact, one can always formulate Einstein-{\ae}ther theory without the constraint term, provided one is careful about implementing the unit norm constraint~\eq{ae:norm} in the variation of the {\ae}ther. In light of this, the dynamics of Ho\v{r}ava (IR limit) and Einstein-{\ae}ther theories can be derived from identical actions.} in the IR limit~\cite{Germani:2009yt, BPS:HL-xm-inconsistency, J:hso-ae=hor}. The key difference is, of course, that in Ho\v{r}ava theory the fundamental field is the scalar field $U$, whereas it is the full vector field $u^a$ in Einstein-{\ae}ther theory. In particular, the equations of motion of Ho\v{r}ava-Lifshitz theory in the IR limit are obtained by extremizing the action~\eq{ae:action} (without the constraint term) with respect to variations of the metric and the scalar field $U$, and this results in a different set of equations than~\eq{ae:EOM}. We will review the equations of Ho\v{r}ava theory in the following section after introducing some formal tools.

For readers who are more familiar/comfortable with Ho\v{r}ava gravity jargons, the following dictionary should help to translate between Einstein-{\ae}ther theory couplings and Ho\v{r}ava-Lifshitz couplings
	\beql{eq:dictionary}
	\xi = \fr{1}{1 - c_{13}}~, \quad \la = \fr{1 + c_2}{1 - c_{13}}~, \quad \eta = \fr{c_{14}}{1 - c_{13}}~.
	\eeq
In particular, the couplings $c_{13}$ and $c_{14}$ of Einstein-{\ae}ther theory are directly related to the couplings $\xi$ and $\eta$ of Ho\v{r}ava-Lifshits theory. A little algebra further reveals
	\beqn{
	c_{123} = \xi^{-1}(\la - 1)~, \qquad c_l = (2\xi)^{-1}(3\la - 1)~.
	}
As well, the overall normalization $G_{\textsc{h}}$ of the Ho\v{r}ava-Lifshitz theory action is related to $\Gae$ according to $G_{\textsc{h}} = \xi\Gae$.

The stability analysis in IR Ho\v{r}ava gravity~\cite{BPS:extHL} is similar to that in Einstein-{\ae}ther theory. In particular, due to the {\ae}ther being hypersurface orthogonal, the propagating modes in Ho\v{r}ava gravity are just the spin-0 and spin-2 modes, with the speeds given by~\eq{speeds}. In terms of the Ho\v{r}ava theory couplings, this translates to
	\beqn{
	s_2^2 = \xi~, \quad s_0^2 = \fr{(\la - 1)\xi}{(3\la - 1)}\lf(\fr{2\xi}{\eta} - 1\rf)~.
	}
In the following, we will almost always express things in terms of the Einstein-{\ae}ther version of the couplings.
\subsection{Comparison of the equations of motion}
In general, an {\ae}ther field in Einstein-{\ae}ther theory satisfying~\eq{ae:norm} and~\eq{ae:EOM} will have a non-vanishing twist, measured by $u \wg \du$. To see how the twist enters the evolution of the {\ae}ther explicitly in Einstein-{\ae}ther theory, let us consider the standard decomposition of the {\ae}ther congruence
	\beqn{
	\Dl_a u_b = -u_a a_b + K_{a b} + w_{a b}~, \quad K_{[a b]} = 0~, \quad w_{(a b)} = 0~,
	}
where $a^a$ is the acceleration of the {\ae}ther defined as below,
	\beqn{
	a^a \equiv \Dl_u u^a~,
	}
the trace and the trace-free parts of $K_{a b}$ are the expansion and shear of the congruence, and the two-form $w_{a b}$ (the vorticity/rotation of the congruence) is directly related to the twist of the {\ae}ther via $u \wg \du = 2(u \wg w)$. The quantities $a^a$, $K_{a b}$, and $w_{a b}$ are all orthogonal to the {\ae}ther by construction. When the above decomposition of $\Dl_a u_b$ is plugged into~\eq{ae:EOM}, the equations of motion of Einstein-{\ae}ther theory can be viewed as first order evolution equations of the acceleration, expansion, shear and the twist.

To understand the relation between the solution spaces of Einstein-{\ae}ther and Ho\v{r}ava-Lifshitz theories, let us focus on the twist-free sector of the solution space of Einstein-{\ae}ther theory. This could be achieved simply by setting $w_{a b} = 0$ globally in the equations of motion~\eq{ae:EOM}, and seeking solutions to the resulting equations. When the equations of motion admit non-trivial twist-free solutions is an interesting open question with important implications\ft{For example, in the context of rotating black hole solutions, twist-free solutions of Einstein-{\ae}ther theory do not exist~\cite{BS:aebh:R1, BS:aebh:R2}.}. Furthermore, as argued in~\cite{J:undo-w}, solutions of Einstein-{\ae}ther theory with a non-vanshing twist should have a non-trivial, well-behaved, twist-free limit when $\bar{c}_{13}$ is taken to infinity, under which these solutions become solutions of Ho\v{r}ava-Lifsitz theory, and therefore, are not necessarily part of the twist-free sector of Einstein-{\ae}ther theory.

These issues are, however, tangential to our main concerns in this paper. In the following, we will enforce the twist-free condition on the {\ae}ther kinematically, by assuming spherical symmetry. Since a solution of Einstein-{\ae}ther theory with zero {\ae}ther twist can only depend on the parameters $c_2$, $c_{13}$ and $c_{14}$~\cite{J:hso-ae=hor}, the process of taking $\bar{c}_{13} \to \infty$ is irrelevant here; all other relevant coefficients will be kept finite and consistent with~\eq{c_i:constraints} in the following. Note that by Frobenius' theorem, an {\ae}ther with a vanishing twist is hypersurface orthogonal and expressible as in~\eq{ae:HSO}.

In Ho\v{r}ava-Lifshitz theory the {\ae}ther is always hypersurface orthogonal. Yet, the difference in the implementations of hypersurface orthogonality on the {\ae}ther in the two theories result in a subtly different set of equations of motions, even though the equations follow from the same action. In particular, the twist-free sector of the Einstein-{\ae}ther theory forms a subset (could be empty) of the solution space of the IR limit of Ho\v{r}ava-Lifshitz theory~\cite{J:hso-ae=hor}. In the remainder of this section, we will review this connection closely following~\cite{J:hso-ae=hor}, to prepare the reader for the proof of the one-to-one equivalence of the spherically symmetric black hole solution spaces.

When the {\ae}ther is hypersurface orthogonal, with the constant $U$ hypersurfaces denoted by $\Sg_U$, the spacetime metric $\met_{a b}$ induces a metric $\pmet_{a b}$ on each member of $\Sg_U$ given by
	\beqn{
	\pmet_{a b} = u_a u_b + \met_{a b}~.
	}
The covariant derivative of the {\ae}ther decomposes as before, but without the twist
	\beql{Du:decomp:HSO}
	\Dl_a u_b = -u_a a_b + K_{a b}~, \qquad a_a = \vecDl_a\log N~.
	\eeq
Note that $K_{a b}$ now admits the interpretation of the extrinsic curvature of the embedding of the preferred foliation into the spacetime, while the acceleration can be expressed as a pure gradient on the hypersurfaces, where $\vecDl_a$ is the spatial covariant derivative compatible with the induced metric $\pmet_{a b}$. The trace of the extrinsic curvature $K_{a b}$ will be denoted by $K$.

The above kinematic considerations are applicable to both Ho\v{r}ava-Lifshitz theory as well as the twist-free sector of Einstein-{\ae}ther theory. In terms of these, the equations of motion for both theories can be compactly presented as follows
	\beql{EEq:HSO}
	\EinG_{a b} = 
	\begin{cases}
	-\displaystyle{\fr{3c_{\CC}}{\ell^2}}\met_{a b} + \aeT_{a b}~, &~\text{Einstein-{\ae}ther}~(w_{a b} = 0)~, \\
	& \\
	-\displaystyle{\fr{3c_{\CC}}{\ell^2}}\met_{a b} + \aeTH_{a b}~, &~\text{Ho\v{r}ava-Lifshitz (IR)}~.
	\end{cases}
	\eeq
The {\ae}ther stress-tensors, respectively in Einstein-{\ae}ther and Ho\v{r}ava-Lifshitz theories, can be decomposed in the {\ae}ther's frame without any loss of generality 
	\beqn{
	\begin{split}
	& \aeT_{a b} = \aeT_{u u}u_a u_b - (u_a\aePi_b + u_b\aePi_a) + \aebT_{a b}~, \\
	& \aeTH_{a b} = \aeT_{u u}u_a u_b - (u_a\PiH_b + u_b\PiH_a) + \aebT_{a b}~.
	\end{split}
	}
In particular, the `$uu$-components' of the {\ae}ther stress tensors $\aeT_{u u}$, as well as the `purely spatial components' $\aebT_{a b}$, are formally identical in both the theories. In terms of the quantities defined in~\eq{Du:decomp:HSO}, the various stress tensor components can then be given as follows. To begin with, the Lagrangian $\lag_{\ae}$, which now has the same form in both theories, reads
	\beql{ae:lag:HSO}
	\lag_{\ae} = -c_{13}K_{ab}K^{ab} - c_2K^2 + c_{14}a^2~.
	\eeq
In terms of the above, $\aeT_{u u}$ is given by
	\beql{Tuu:HSO:ae}
	\aeT_{u u} = \fr{\lag_{\ae}}{2} + c_{14}(\vecDl\cdot a)~,
	\eeq
while the purely spatial components are
	\beql{Tij:HSO:ae}
	\begin{split}
	\aebT_{a b} = \{c_2\Dl_c[Ku^c] & + (1/2)\lag_{\ae}\}\pmet_{a b} - c_{14}a_a a_b~\\
	& + c_{13}[KK_{a b} + \tn{\pmet}{_a^c}\tn{\pmet}{_b^d}(\Dl_uK_{c d})]~.
	\end{split}
	\eeq
On the other hand, the `cross-components' of the stress tensors, $\aePi_a$ for Einstein-{\ae}ther theory and $\PiH_a$ for Ho\v{r}ava-Lifshitz theory, have different structures\ft{If one treats the covector form of the {\ae}ther field as the fundamental field in Einstein-{\ae}ther theory, then the `cross-component' of the {\ae}ther stress tensor is identical with $\PiH_a$ in the twist-free sector. On-shell, both descriptions are equivalent due to~\eq{reln:PiPi-AE}.}. In particular, for Ho\v{r}ava-Lifshitz theory, we have
	\beql{PiH:HSO:ae}
	\PiH_a = c_{13}(\vecDl_c\tn{K}{^c_a}) + c_2\vecDl_a K~,
	\eeq
whereas for the twist-free sector of Einstein-{\ae}ther theory, we have
	\beql{aePi:HSO:ae}
	\aePi_a = c_{14}[Ka_a + (\Dl_u a^c)\pmet_{c a} - K_{a c}a^c]~.
	\eeq
The vectors $\aePi_a$ and $\PiH_a$ are furthermore related by
	\beql{reln:PiPi-AE}
	\vecAE_a \equiv \tn{\pmet}{_a^b}\AE_b = \PiH_a - \aePi_a~,
	\eeq
where, to remind the reader, $\AE_a$ is the functional derivative of the {\ae}ther action with respect to $u^a$ as introduced in~\eq{ae:EOM}. Note that the relation~\eq{reln:PiPi-AE} is kinematical and hence valid off-shell.

Now, the {\ae}ther's equations of motion in the twist-free sector of Einstein-{\ae}ther theory is just the twist free version of the second equation in~\eq{ae:EOM}
	\beql{ae:EOM:HSO}
	\vecAE_a = 0~.
	\eeq
On the other hand, upon extremizing the {\ae}ther action under variations of the scalar field $U$~\eq{ae:HSO}, the {\ae}ther's equation of motion in Ho\v{r}ava-Lifshitz theory turns out to be
	\beql{HL:ae:EOM}
	\Dl_a[N\vecAE^a] = 0~.
	\eeq
Manifestly,~\eq{ae:EOM:HSO} is different from~\eq{HL:ae:EOM}, and the difference clearly stems from imposing the hypersurface orthogonality condition on the {\ae}ther at the level of the action (Ho\v{r}ava-Lifshitz theory) versus at the level of the equations of motion (Einstein-{\ae}ther theory).

Yet on the other hand, an {\ae}ther-metric configuration (with a hypersurface orthogonal {\ae}ther) that satisfies~\eq{ae:EOM:HSO} and the Einstein's equations for Einstein-{\ae}ther theory~\eq{EEq:HSO}, not only trivially satisfies~\eq{HL:ae:EOM}, but also the Einstein's equations of Ho\v{r}ava-Lifshitz theory~\eq{EEq:HSO}, owing to the identity~\eq{reln:PiPi-AE}. The converse of this statement is clearly not true in general -- an {\ae}ther-metric configuration satisfying~\eq{HL:ae:EOM} may not satisfy~\eq{ae:EOM:HSO}. In other words, the twist-free sector of Einstein-{\ae}ther theory forms a (proper) subset of the solution space of Ho\v{r}ava-Lifshitz theory~\cite{J:hso-ae=hor}.

In section~\ref{ss:equiv}, we will prove that these sets are in fact identical in the case of static and spherically symmetric black hole solutions. To that end, let us briefly review the implementation of these symmetries in the present context (see~\cite{JB:thesis} for details).
\subsection{Spherical symmetry \& staticity: kinematics}\label{background:ss}
In a spherically symmetric situation, the {\ae}ther is twist-free kinematically. One may furthermore define $s^a$ to be the unit spacelike vector along the acceleration (and therefore orthogonal to the {\ae}ther), i.e.,
	\beqn{
	a^a = (a\cdot s) s^a~,~~~(u\cdot s) = 0~,~~~s^2 = 1~,~~~(a\cdot s)^2 = a^2~.
	}
Any rank-two tensor can then be expanded in a basis spanned by the bi-vectors $u_a u_b$, $u_{(a} s_{b)}$, $u_{[a} s_{b]} \equiv (1/2)\veii_{a b}$, $s_a s_b$ and $\hatmet_{a b}$ (the projector/induced metric on the two spheres). For example, 
	\beqn{
	\Dl_a u_b = -(a\cdot s)u_a s_b + K_{a b}, \quad K_{a b} = K_{s s} s_a s_b + \fr{\hatK}{2}\hatmet_{a b}~,
	}
and likewise
	\beqn{
	\Dl_a s_b = -(a\cdot s) u_a u_b + K_{s s} s_a u_b + \fr{\hatk}{2}\hatmet_{a b}~.
	}
The reader is encouraged to consult the relevant parts of~\cite{BBM:mechuhor} or~\cite{JB:thesis} for further information about these equations, if the need be. As will be shown below, $(a\cdot s)$, $K_{s s}$, $\hat{K}$ and $\hat{k}$ can be computed once a particular coordinate system has been chosen and the metric is given in that coordinate basis. Taking the trace of the first relation, the mean curvature $K$ works out to be
	\beqn{
	K = K_{s s} + \hat{K}~.
	}
The various stress tensor components now simplify as follows: the `$uu$-component'~\eq{Tuu:HSO:ae} becomes
	\beqn{
	\aeT_{u u} = \fr{\lag_{\ae}}{2} + c_{14}\lf[\Dl_s(a\cdot s) + \hatk(a\cdot s)\rf]~,
	}
while the purely spatial part~\eq{Tij:HSO:ae} simplifies to
	\beqn{
	\aebT_{a b} = \aeT_{s s}s_a s_b + \fr{\hat{\aeT}}{2}\hatmet_{a b}~,
	}
where
	\beqn{
	\aeT_{s s} = c_{13}\Dl_c[K_{s s} u^c] + c_2\Dl_c[K u^c] - c_{14}(a\cdot s)^2 + \fr{\lag_{\ae}}{2}~,
	}
and
	\beqn{
	\hat{\aeT} = c_{13}\Dl_c[\hatK u^c] + 2c_2\Dl_c[K u^c] + \lag_{\ae}~.
	}
Furthermore, both $\aePi_a$~\eq{aePi:HSO:ae} and $\PiH_a$~\eq{PiH:HSO:ae} are directed along $s_a$ and given by [$\PiH_s \equiv (s\cdot\PiH)$]
	\beqn{
	\PiH_a = \PiH_s s_a~, \quad \PiH_s = c_{13}[\Dl_sK_{s s} + (K_{s s} - (\hatK/2))\hatk] + c_2\Dl_sK~,
	}
and [$\aePi_s \equiv (s\cdot\aePi)$]
	\beqn{
	\aePi_a = \aePi_s s_a~, \quad \aePi_s = c_{14}[\Dl_u(a\cdot s) + \hatK(a\cdot s)]~.
	}
One can then find $\vecAE_a$ from~\eq{reln:PiPi-AE}, which also has a component only along $s_a$.

So far we have only explored the consequences of spherical symmetry (therefore, in particular, the above expressions can be used to study time dependent spherically symmetric dynamics of Einstein-{\ae}ther/Ho\v{r}ava-Lifshitz theories). However, with our goal of studying black holes, let us further restrict the kinematics to static backgrounds, characterized by the existence of a time translation generating Killing vector $\chi^a$. Spherical symmetry dictates that $\chi^a$ can be expanded in the $\{u^a, s^a\}$ basis as
	\beql{def:X^a}
	\chi^a = -(u\cdot\chi)u^a + (s\cdot\chi)s^a~, \qquad \Dl_a \chi_b = -\ka\,\veii_{a b}~,
	\eeq
where the `surface-gravity' $\ka$ is\ft{Strictly speaking, the surface gravity is only defined on a Killing horizon, while $\ka$, as defined in~\eq{def:ka}, is the redshifted acceleration of the static observer as measured at infinity. We will still call $\ka$ the surface gravity at a given radial location by a (mild) abuse of terminology.}
	\beql{def:ka}
	\ka = -(a\cdot s)(u\cdot\chi) + K_{s s}(s\cdot\chi)~.
	\eeq
Any general vector respecting spherical symmetry can be decomposed along $u^a$ and $s^a$ as in the first relation of~\eq{def:X^a}. Making $\chi^a$ a Killing vector, i.e., making it satisfy Killing's equation, which is equivalent to the second relation in~\eq{def:X^a} amounts to further requiring~\eq{def:ka} and
	\beql{reln:hatk-hatK}
	(u\cdot\chi)\hat{K}= (s\cdot\chi)\hat{k}~.
	\eeq
In the following, we will work in the ingoing Eddington-Finklestein (EF) coordinate system, which is well defined everywhere except at $r = 0$, to express the metric and the {\ae}ther components. In these coordinates, the metric takes the form (all the coordinates have their usual meaning)
	\beql{met:EF}
	\ds^2 = -e(r)\dv^2 + 2f(r)\dv\dr + r^2\dOm_2^2~,
	\eeq
where $\dOm_2^2 = (\dtH^2 + \sin^2\tH\dphi^2)$ is the metric on the two-spheres. The timelike Killing vector is $\chi^a = \dl_v$ in these coordinates. Note that the metric component $e(r) = -\chi^2$ is the negative of the norm of the Killing vector. It will also be useful to introduce the radial vector $\rho^a$ and the corresponding one form $\rho_a$, with the latter taking a particularly simple form here
	\beql{def:rho}
	\rho_a = f(r)\Dl_ar~.
	\eeq
One may verify that the radial vector is everywhere well defined, is orthogonal to the Killing vector $\chi^a$ and has the opposite norm of it, i.e., $(\rho\cdot\chi) = 0$ and $\rho^2 + \chi^2 = 0$. Furthermore, the {\ae}ther can be expressed in the present coordinate system as
	\beql{u:EF}
	u_a = (u\cdot\chi)\dv + \fr{f(r)\dr}{(s\cdot\chi) - (u\cdot\chi)}~, \quad e(r) = (u\cdot\chi)^2 - (s\cdot\chi)^2~,
	\eeq
where the second relation takes care of the unit norm constraint~\eq{ae:norm}, while the orientations are fixed by demanding $(u\cdot\rho) = -(s\cdot\chi)$ and $(s\cdot\rho) = -(u\cdot\chi)$.

To address causality in such backgrounds, we further need to understand the behaviour of the constant {\ae}ther slices (as the metric/light cones no longer determine causality). This could be achieved, for instance, by studying the behaviour of the {\ae}ther time function $U$~\eq{ae:HSO} in the present coordinate system. However, $U$ itself can only be determined up to the reparametrizations as in~\eq{U:reparams}. Interestingly, even though the lapse function $N$ also depends on the choice of $U$, it does so in a `separable' manner as follows~\cite{BCS:defUH}
	\beql{reln:N-u.X:gen}
	N = F(U)(u\cdot\chi)~,
	\eeq
where $F(U)$ is some function of $U$. One possible way to fix the above arbitrariness is to demand that $U$ satisfies some extra condition, e.g., $\chi^a\Dl_aU = 1$; imposing this choice in~\eq{ae:HSO} yields a `gauge-fixed' preferred frame lapse function
	\beql{reln:N-u.X}
	N = -(u\cdot\chi)~,
	\eeq
and for this choice of gauge, the {\ae}ther time function $U$ can be explicitly read off from~\eq{u:EF}
	\beqn{
	u_a = (u\cdot\chi)\dU, \quad \dU = \dv + \fr{f(r)\dr}{(u\cdot\chi)[(s\cdot\chi) - (u\cdot\chi)]}~.
	}
Although it won't be directly relevant, we may record the form of the metric in the $\{U, r\}$ coordinates adapted to the preferred {\ae}ther hypersurfaces
	\beqn{
	\ds^2 = -(u\cdot\chi)^2\dU^2 + \lf[(s\cdot\chi)\dU - \fr{f(r)\dr}{(u\cdot\chi)}\rf]^2 + r^2\dOm_2^2~.
	}
In these coordinates, $\chi^a = \dl_U$ and hence the shift vector is simply $(s\cdot\chi)s^a$.
	\begin{figure}
	\centering
	\includegraphics[scale=0.9]{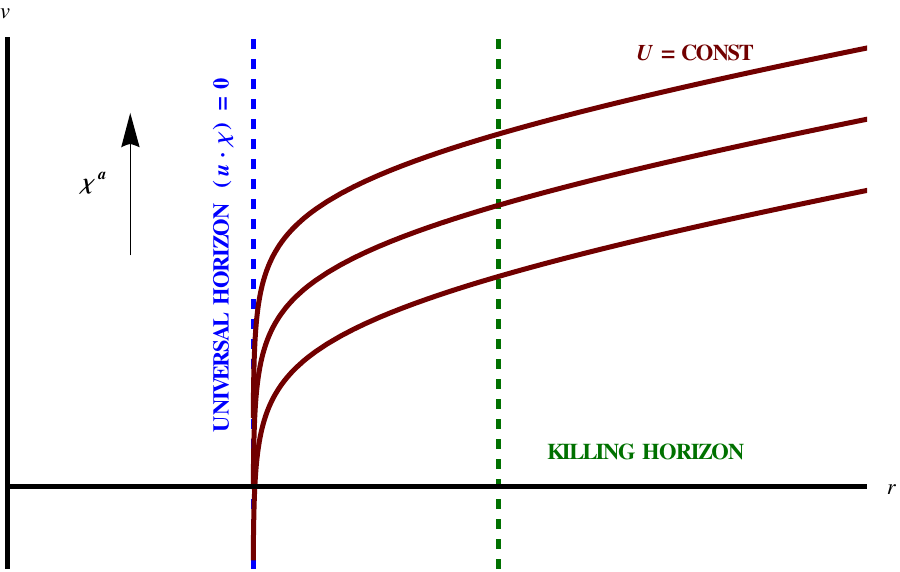}
	\caption{Bending of the \warnbr{$\Sg_U$ hypersurfaces} (\warnbr{\bf thick brown lines}) near the \warnb{universal horizon} (\warnb{\bf blue dotted line}) of an {\ae}ther black hole. The \warng{\bf green dotted line} denotes the usual \warng{Killing horizon} ($\chi^2 = 0$). The Killing vector $\chi^a$ points upward throughout everywhere. The constant {\ae}ther hypersurfaces inside the universal horizon (i.e., for $r < \ruh$) has not been shown to keep the diagram clean.}
	\label{fig:bendingU}
	\end{figure}

One may now consider the behaviour of $U = $~constant hypersurfaces in the $v$-$r$ plane (figure~\ref{fig:bendingU} captures a `typical behaviour'). As shown in figure~\ref{fig:bendingU}, each $U = $~constant hypersurface bends down to the infinite past in $v$ as one moves towards $r = 0$ and asymptotes to a 3 dimensional spacelike hypersurface defined by $(u\cdot\chi) = 0$.

We can turn this surface into a causal boundary by requiring that the matter field equations are such that matter excitations always travel to the future in $U$ time. This is equivalent to the requirement that $\emT^{\mathrm{matter}}_{u u} \geqq 0$, i.e., the {\ae}ther version of the weak energy condition on the matter stress tensor. Note that this condition still allows for arbitrarily fast excitations in the {\ae}ther frame, in agreement with the non-relativistic symmetry of the background. With this restriction, signals propagating to the future in $U$ must necessarily move towards decreasing $r$ from this surface. Therefore, the three-dimensional hypersurface
	\beqn{
	(u\cdot\chi) = 0~,
	}
defines a {\it universal horizon}, as it traps even arbitrarily fast excitations. Note that the metric in preferred coordinate has a coordinate singularity at the universal horizon; this is to be exprected due to the behaviour of the preferred time coordinate as the universal horizon is approached.

If the equation $(u\cdot\chi) = 0$ has more than one root, there could be multiple universal horizons, one nested inside another. When this is the case, the outermost one plays the role of the causal boundary. We will denote the radius of the cross section of the universal horizon by $\ruh$ (so that the universal horizon is `located' at $r = \ruh$). Due to spherical symmetry, $s^a$ becomes the unit vector along $\chi^a$ on the universal horizon and $(s\cdot\chi)_{\uhor} = \Xuh$ where $\Xuh$ is the magnitude of $\chi^a$ on the universal horizon.

Our coordinate choice and the parametrization of the {\ae}ther field in these coordinates show that in a static and spherically symmetric situation, there are three independent functions that completely specify the geometry and the {\ae}ther field configuration, namely $(u\cdot\chi)$, $(s\cdot\chi)$ and $f(r)$. All other quantities can be derived from these three functions and their derivatives; e.g., the coefficients in the basis-expansion of $\Dl_a u_b$ and $\Dl_a s_b$ are\ft{Given any scalar function $F(r)$, the symmetries imply $\Dl_aF(r) = f(r)^{-1}F'(r)\rho_a$. In particular, this means $\Dl_uF(r) = -(s\cdot\chi)f(r)^{-1}F'(r)$ and $\Dl_sF(r) = -(u\cdot\chi)f(r)^{-1}F'(r)$, using the relations following~\eq{def:rho}.}
	\beql{sss:misc-expr}
	\begin{split}
	& (a\cdot s) = -\fr{(u\cdot\chi)'}{f(r)}, \qquad K_{s s} = -\fr{(s\cdot\chi)'}{f(r)},~\\
	& \hat{K} = -\fr{2(s\cdot\chi)}{rf(r)}, \qquad\qquad \hat{k} = -\fr{2(u\cdot\chi)}{rf(r)}~,
	\end{split}
	\eeq
where a prime ($\prime$) denotes a derivative with respect to $r$. One may now plug these expressions directly into the equations of motion of Einstein-{\ae}ther~\eq{EEq:HSO},~\eq{ae:EOM:HSO} and/or Ho\v{r}ava-Lifshitz~\eq{EEq:HSO},~\eq{HL:ae:EOM} theory and look for solutions, as we do in the next section.
\section{Spherically symmetric solutions}
In the previous section, we introduced the notion of the universal horizon and explained how it acts as a causal/event horizon even for arbitrarily fast excitations. An {\ae}ther-metric background with a universal horizon is therefore a black hole. In this section, we will first prove the equivalence of static and spherically symmetric {\ae}ther black hole solution spaces in Ho\v{r}ava-Lifshitz and Einstein-{\ae}ther theories. Subsequently, we will classify all static and spherically symmetric {\ae}ther black hole solutions in these theories with maximally symmetric asymptotics.
\subsection{The equivalence of black hole solution spaces}\label{ss:equiv}
We already know that static and spherically symmetric solutions of Einstein-{\ae}ther theory will be a subset of the solutions of Ho\v{r}ava-Lifshitz theory~\cite{J:hso-ae=hor}. To show equivalence, we therefore must prove the converse. Substituting the `separable form'~\eq{reln:N-u.X:gen} of the lapse function $N$ into the equation of motion of the {\ae}ther~\eq{HL:ae:EOM} turns the latter into
	 \beqn{
	 F(U)\Dl_a[(u\cdot\chi)\vec{\AE}^a] + (u\cdot\chi)\vec{\AE}^a\Dl_aF(U) = 0~.
	 }
Now, the last term vanishes since $\vec{\AE}^a\Dl_a$ is a `spatial derivative' along the {\ae}ther hypersurfaces. We are therefore left with a manifestly reparametrization invariant form of the equation of motion for the {\ae}ther in Ho\v{r}ava-Lifshitz theory
	\beql{HL:ae:equiv}
	\Dl_a[(u\cdot\chi)\vec{\AE}^a] = 0~.
	\eeq
This equation is now particularly easy to integrate with the additional assumptions of staticity and spherical symmetry. Once this is done, the result can be expressed as follows
	\beqn{
	-(u\cdot\chi)(\rho\cdot\vecAE) = (u\cdot\chi)^2(s\cdot\vecAE) = \fr{C_0}{r^2}~,
	}
where, we have used properties of the radial vector $\rho_a$, discussed in the paragraph following~\eq{def:rho}, as well as introduced $C_0$ as the constant of integration. Note that the above expressions are entirely in terms of coordinate independent quantities, except for the explicit appearance of $r^2$ after the second equality.

Now, if $C_0 \neq 0$, then the above expression requires $(s\cdot\vecAE)$ to diverge whenever $(u\cdot\chi)$ vanishes -- i.e., at the universal horizon of an {\ae}ther black hole -- making the solution singular there. Therefore demanding regularity of a black hole solution in the bulk of the spacetime between the universal horizon and infinity forces $C_0 = 0$. However, by spherical symmetry this means $\vecAE_a = 0$, so that the {\ae}ther configuration satisfies the {\ae}ther's equation of motion~\eq{ae:EOM:HSO} in Einstein-{\ae}ther theory. Furthermore, from the equality of the stress tensor components (particularly, due to~\eq{reln:PiPi-AE}), the background also satisfies the relevant Einstein's equations of Einstein-{\ae}ther theory. In other words, {\it every regular static and spherically symmetric {\ae}ther black hole solution of Ho\v{r}ava-Lifshitz theory is also a solution of Einstein-{\ae}ther theory}. Hence the solution spaces are equivalent.

As should be clear from our arguments above, the asymptotic behaviour of the solutions play no role in our proof of the equivalence of the solution spaces. In addition, the presence of matter does not affect the functional forms of the above expressions -- hence the proof holds whether or not matter is present. The only role played by spherical symmetry is to make the problem `effectively two dimensional', so the proof additionally holds for transverse spaces with other geometries (e.g., flat $\mathbf{R}^2$, commonly studied in the context of Lifshitz solutions). Finally, one could formulate both Einstein-{\ae}ther and Ho\v{r}ava-Lifshitz theories in an arbitrary number of spacetime dimensions, and the equations of motion essentially stay intact; in particular, the {\ae}ther's equations of motion are identical to~\eq{ae:EOM:HSO} and~\eq{HL:ae:EOM} respectively. It should be obvious that the above argument naturally generalizes to arbitrary dimensions as well. Therefore, our conclusions in this section are applicable to much more general solutions than the black hole solutions that we are going to present next, and somewhat more general than previous proofs about the solution space equivalence.
\subsection{Boundary conditions and simplifying the equations of motion}\label{sss:aebh}
We are looking for solutions with universal horizons with maximally symmetric asymptotics that are smoothly connected to globally maximally symmetric solutions. By maximally symmetric asymptotics we mean that the metric components~\eq{met:EF} have the following behavior at large $r$
	\beql{def:BC}
	e(r) \sim -\fr{\La r^2}{3} + 1 + \ord(r^{-1})~, \quad f(r) \sim 1 + \ord(r^{-1})~,
	\eeq
where $\La$, with dimensions length$^{-2}$, is the `effective' cosmological constant of the background. As we will see below, $\La$ may differ from the bare cosmological constant that appears in the action~\eq{ae:action}. Solutions with (negative) positive $\La$ are asymptotically (anti-)de Sitter, while for $\La = 0$, we have asymptotically flat spacetime. We do not impose any corresponding restriction on the behavior of the {\ae}ther profile at infinity, except for that imposed by the second relation in~\eq{u:EF} due to the asymptotic behaviour of $e(r)$ as prescribed in~\eq{def:BC}.

Given the above boundary conditions, we can insert the leading order behavior of the metric components~\eq{def:BC} in the relevant equations of motion and derive all possible {\ae}ther profiles for globally maximally symmetric solutions (assuming one can perform such an analysis in practice). These cases will, furthermore, cover all possible allowed behavior of the {\ae}ther at infinity, consistent with our assumptions.

As it turns out, we can actually do better: we can, in fact, give up any restriction on $e(r)$ as well, and carry out a complete analysis of the equations of motion only with setting $f(r) = 1$. Of course, we will also be interested in more general black hole solutions which have $f(r) \neq 1$ in the bulk. However, as we will elaborate more later, any such solution must be continuously connected to some appropriate `{\ae}ther-vacuum' solution with the latter having $f(r) = 1$ globally. In fact, since the analysis with $f(r) = 1$ does not assume any restriction on the behaviour of $e(r)$, even asymptotically, such an analysis could have potentially captured spacetimes with a more general asymptotic behavior than that assumed in~\eq{def:BC}; we will however find that no such solution exists. Therefore, finding all solutions with $f(r) = 1$ will necessarily yield {\it all} allowed maximally symmetric vacuum solutions of Einstein-{\ae}ther and Ho\v{r}ava-Lifshitz theory.

We start by noting that for a general static and spherically symmetric background the `cross-component' $\Ric_{u s} \equiv \Ric_{a b}u^a s^b$ has the following simple form
	\beqn{
	\Ric_{u s} = \fr{2(s\cdot\chi)(u\cdot\chi)f'(r)}{rf(r)^3}~.
	}
Therefore, when $f(r) = 1$, $\Ric_{u s} = 0$ (see also~\cite{J:gtt-grr}). One may then consider appropriate linear combinations of the Einstein's equation $\Ric_{u s} = \aePi_s$~\eq{EEq:HSO} and the {\ae}ther's equation of motion~\eq{ae:EOM:HSO} to conclude the following equivalence
	\beql{sss:central}
	\Ric_{u s} = 0 \quad\iim\quad \aePi_s = 0 \quad\iim\quad c_{123}\Dl_sK = 0~.
	\eeq
Furthermore, when $f(r) = 1$, the general expressions for $\aePi_s$ and $\Dl_sK$ simplify, upon using~\eq{sss:misc-expr}, to
	\beqn{
	\begin{split}
	& \aePi_s = \fr{c_{14}(s\cdot\chi)}{r^2}\lf[r^2(u\cdot \chi)'\rf]'~,~\\
	& \Dl_s K = (u\cdot\chi)\lf[r^{-2}[r^2(s\cdot\chi)]'\rf]'~,
	\end{split}
	}
where as before, a prime denotes a derivative with respect to $r$. Therefore, there are only finitely many ways to satisfy~\eq{sss:central}, and one needs to consider these possibilities on a case-by-case basis; the curious reader is encouraged to consult section 2.2.1 of~\cite{JB:thesis} for further analysis. Note that some of the above equations may be solved trivially for either $c_{14} = 0$ or $c_{123} = 0$. Furthermore, these two `special points' in the coupling space are physically interesting for the following reason: the spin-0 speed $s_0$~\eq{speeds} diverges as $c_{14} \to 0$, while it vanishes as $c_{123} \to 0$\ft{Of course, there are other limits of~\eq{speeds} when $s_0$ can vanish or diverge: $c_{14} \to 2$ ($s_0$ vanishes), $c_{13} \to 1$ ($s_0$ diverges) and $c_l \to 0$ ($s_0$ diverges). However, they all violate the constraints \eq{c_i:constraints}, and therefore are excluded.}.

With $f(r)$ set to unity and the resulting simplifications that follow, one is led to consider the following three mutually exclusive\ft{Setting both $c_{123}$ and $c_{14}$ to zero simultaneously is unphysical, as it removes the {\ae}ther's equation of motion (equivalently, the evolution equation of the lapse $N$ in the `Ho\v{r}ava picture') from the available set of equations, and one is simply left with the vacuum Einstein's equations with a preferred yet non-dynamical {\ae}ther.} cases
	\begin{enumerate}
	\item the case corresponding to generic values of couplings ($c_{123} \neq 0$, $c_{14} \neq 0$) consistent with~\eq{c_i:constraints},
	\item the special case corresponding to $c_{123} = 0$, and
	\item the special case corresponding to $c_{14} = 0$.
	\end{enumerate}
Further analyses of the remaining Einstein's equations then yield either globally Minkowski or de Sitter solutions for generic values of the coupling, or non-trivial analytic solutions, with all kinds of maximally symmetric asymptotics, parametrized by more than one integration constants for both the cases $c_{123} = 0$ and $c_{14} = 0$. When the `mass parameter' of the solutions for these special coupling cases is `turned off'\ft{The complete solutions for these special coupling cases in their full generality are shown in equations~\eq{c123=0} and~\eq{c14=0} below.}, one ends up with the corresponding globally maximally symmetric solutions.
	\begin{table*}[t]
	\centering
	\begin{tabular}{||c|c|c|c||}
	\hline
	                 &                                                                       &                                                                       & \\
	globally         & generic couplings ($c_{123}, c_{14} \neq 0$)                          & $c_{123} = 0$                                                         & $c_{14} = 0$ \\
	                 &                                                                       &                                                                       & \\
	\hline \hline
	                 &                                                                       &                                                                       & \\
	                 &~~$(u\cdot\chi) = -1$~,~$(s\cdot\chi) = \displaystyle{\fr{r}{\elef}}$~~&~~$(u\cdot\chi) = -1$~,~$(s\cdot\chi) = \displaystyle{\fr{r}{\elef}}$~~&~~$(u\cdot\chi) = -\rt{\displaystyle{\fr{r^2}{\ell_u^2}} + 1}$~,~$(s\cdot\chi) =\displaystyle{\fr{r}{\ell_s}}$~~\\
	                 &                                                                       &                                                                       & \\
        de Sitter        & \quad $\elef = \ell\rt{c_l}$                                          &~$\elef = \ell\rt{1 - c_{13}}$                                         & $\ell_u^2 = \ell^2\ell_s^2(c_l\ell^2 - c_{\CC}\ell_s^2)^{-1}$,~~$\ell_u > \ell_s$ \\
	                 &                                                                       &                                                                       & \\              
                         & Free parameters: none.                                                & Free parameters: none.                                                & Free parameters: $\ell_s$~. \\
	                 & Corresponding black hole:~\eq{generic-coupling:dS}.                   & Corresponding black hole:~\eq{c123=0}.                                & Corresponding black hole:~\eq{c14=0}. \\
	                 &                                                                       &                                                                       & \\
	\hline \hline
	                 &                                                                       &                                                                       & \\
	                 & $(u\cdot\chi) = -1$~,~$(s\cdot\chi) = 0$~,                            & $(u\cdot\chi) = -1$~,~$(s\cdot\chi) = 0$~,                            &~~$(u\cdot\chi) = -\rt{\displaystyle{\fr{r^2}{\ell_s^2}} + 1}$~,~$(s\cdot\chi) =\displaystyle{\fr{r}{\ell_s}}$~~\\
	                 &                                                                       &                                                                       & \\
	Minkowski        &                                                                       &                                                                       & $c_{\CC}\ell_s^2 = (c_l - 1)\ell^2$~\\
	                 &                                                                       &                                                                       & \\            
                         & Free parameters: none.                                                & Free parameters: none.                                                & Free parameters: $\ell_s$ only if $c_{\CC} = 0$ \\
                         &                                                                       &                                                                       & and $c_l = 1$, otherwise none.\\
	                 & Corresponding black hole:~\eq{generic-coupling:flat}.                 & Corresponding black hole:~\eq{c123=0}.                                & Corresponding black hole:~\eq{c14=0}. \\
	                 &                                                                       &                                                                       & \\
	\hline \hline
	                 &                                                                       &                                                                       & \\
	                 &                                                                       &                                                                       &~~$(u\cdot\chi) = -\rt{\displaystyle{\fr{r^2}{\ell_u^2}} + 1}$~,~$(s\cdot\chi) =\displaystyle{\fr{r}{\ell_s}}$~~\\
	                 &                                                                       &                                                                       & \\
	~anti-de~Sitter~ & No solution.                                                          & No solution.                                                          & $\ell_u^2 = \ell^2\ell_s^2(c_l\ell^2 - c_{\CC}\ell_s^2)^{-1}$,~~$\ell_u < \ell_s$ \\
	                 &                                                                       &                                                                       & \\            
                         &                                                                       &                                                                       & Free parameters: $\ell_s$~. \\
	                 &                                                                       &                                                                       & Corresponding black hole:~\eq{c14=0}. \\
	                 &                                                                       &                                                                       & \\
	\hline
	\end{tabular}
	\caption{Globally maximally symmetric solutions. For all solutions the metric component $f(r)=1$, while $e(r)$~\eq{u:EF} has the generic form $e(r) = -(\La/3)r^2 + 1$ with $\La$ being the effective cosmological constant. Specifically, for the generic coupling and $c_{123} = 0$ cases, $\La = 3\elef^{-2}$ for globally de Sitter solutions~\eq{c14!=0:dS:leff}, while $\La = 0$ for globally Minkowski solutions. When $c_{14} = 0$, $(\La/3) = \ell_s^{-2} - \ell_u^{-2}$~\eq{c14=0:scales}, so that the asymptotic nature of the spacetime depends on the relative sizes of $\ell_u$ and $\ell_s$~\eq{c14=0:asymp}. The global solutions are discussed in more details in the main text; in particular, see~\eq{c14!=0:global:dS},~\eq{c14!=0:global:flat} and~\eq{c14=0:global}. The reference to the general black hole solutions (with a universal horizon) corresponding to each case can be found at the bottom of the respective cells. The number of free parameters is the number of parameters needed to describe the solution given fixed coefficients in the Lagrangian.}
	\label{table:sss:global}
	\end{table*}
\subsection{Globally maximimally symmetric solutions} 
Before we begin the black hole classification, we first determine the global maximally symmetric solutions of Einstein-{\ae}ther theory, as these are required to understand the dynamics of the black hole solutions. The black hole solutions will be smoothly connected to one of the maximally symmetric solutions below, which means that the maximally symmetric solution spaces are also equivalent in the two theories under consideration, due to the result in section~\ref{ss:equiv}. Therefore we only need to focus on Einstein-{\ae}ther theory in what follows. Table~\ref{table:sss:global} lists all such globally maximally symmetric solutions that the equations of motion admit, obtained as described above. 
\subsubsection{Generic coupling case and $c_{123}=0$}
\paragraph{Globally de Sitter:} The globally de Sitter solutions, for both the generic coupling case as well as the case with $c_{123} = 0$, are described by
	\beql{c14!=0:global:dS}
	\begin{split}
	& (u\cdot\chi) = -1~,~~~~(s\cdot\chi) = \fr{r}{\elef}~,~~~~f(r) = 1~, \\
	& e(r) = \lf(1 - \fr{r^2}{\elef^2}\rf)~,
	\end{split}
	\eeq
where, in terms of the bare cosmological constant scale $\ell$, the scale $\elef$ of the effective cosmological constant, which determines the asymptotic behavior of spacetime, is
	\beql{c14!=0:dS:leff}
	\elef =
	\begin{cases}
	\ell\rt{c_l}~, & \text{generic couplings}~,\\
	\ell\rt{1 - c_{13}}~, & c_{123} = 0~.
	\end{cases}
	\eeq
Therefore, the latter class of solutions can be thought of as a special case of the former class, since $c_l = 1 - c_{13}$ when $c_{123} = 0$. One may also note that these solutions have constant mean curvature on the {\ae}ther hypersurfaces given by $K = -3\elef^{-1}$; this is perhaps not surprising given the very high degree of symmetry of the background. Additionally, since $(u \cdot\chi) = -1$ always for globally de Sitter, there is a cosmological Killing horizon but no universal horizon.
\paragraph{Globally Minkowski:} The global Minkowski solutions are also identical for the generic coupling and the $c_{123} = 0$ cases, with {\ae}ther aligned with the Killing vector $\chi^a$. The solutions are given by
	\beql{c14!=0:global:flat}
	(u\cdot\chi) = -1~,~~~(s\cdot\chi) = 0~,~~~f(r) = 1~,~~~e(r) = 1~.
	\eeq
These solutions all have zero mean curvature, i.e., $K = 0$.
\paragraph{Globally anti-de Sitter:} No global anti-de Sitter solutions exist either for the general coupling case or for $c_{123} = 0$.
\subsubsection{$c_{14}=0$}
 Since no global anti-de Sitter solution exists unless $c_{14}=0$, no asymptotically anti-de Sitter solution (characterized by more parameters, e.g., a mass) exists except for this special choice of the couplings either, as they cannot be smoothly taken to a global anti-de Sitter solution of Einstein-{\ae}ther or Ho\v{r}ava-Lifshitz theory. The globally maximally symmetric solutions corresponding to $c_{14} = 0$, and encompassing all kinds of asymptotics -- de Sitter, Minkowski, and anti-de Sitter -- can be compactly presented as
	\beql{c14=0:global}
	\begin{split}
	& (u\cdot\chi) = -\rt{\fr{r^2}{\ell_u^2} + 1}~,~~~(s\cdot\chi) = \fr{r}{\ell_s}~,~~~f(r) = 1~, \\
	& e(r) = -\fr{\La r^2}{3} + 1~,
	\end{split}
	\eeq
where, the constants $\ell_u$, $\ell_s$ and $\La$ are related to each other as well as to parameters of the Lagrangian~\eq{ae:action} through
	\beql{c14=0:scales}
	\begin{split}
	& \fr{c_{\CC}}{\ell^2} = \fr{c_l}{\ell_s^2} - \fr{1}{\ell_u^2} \\
	&~\fr{\La}{3} = \fr{1}{\ell_s^2} - \fr{1}{\ell_u^2}
	\end{split}
	~~~\iim~~~
	\begin{split}
	& \fr{c_l - 1}{\ell_u^2} = \fr{c_{\CC}}{\ell^2} - \fr{c_l\La}{3} \\
	& \fr{c_l - 1}{\ell_s^2} = \fr{c_{\CC}}{\ell^2} - \fr{\La}{3}~.
	\end{split}
	\eeq
In particular, the nature of the solutions are determined according to the sign of the effective cosmological constant $\La$ (recall~\eq{def:BC}) as follows
	\beql{c14=0:asymp}
	\begin{split}
	& \La > 0 \quad\iim\quad \ell_u > \ell_s \quad\iim\quad~\text{de Sitter}~, \\
	& \La = 0 \quad\iim\quad \ell_u = \ell_s \quad\iim\quad~\text{Minkowski}~, \\
	& \La < 0 \quad\iim\quad \ell_u < \ell_s \quad\iim\quad~\text{anti-de Sitter}~.
	\end{split}
	\eeq
Note that $\La$ can have any sign irrespective of the sign of the bare cosmological constant\ft{For example, global Minkowski solutions can arise for $\ell_u^2 = c_{\CC}(c_l - 1)\ell^2$ if $c_{\CC} \neq 0$ and $c_l \neq 1$, or for arbitrary $\ell_u$ (including the case $\ell_u \to \infty$) if $c_{\CC} = 0$ and $c_l = 1$. The corresponding relations for $\La > 0$ or $\La < 0$ can be worked out likewise~\cite{JB:thesis}.}.

Therefore, all global solutions for $c_{14} = 0$ (unlike the global solutions for the generic coupling or $c_{123} = 0$ cases) have generically four length scales associated with them, which are related by~\eq{c14=0:scales}. They are $\ell_u$ and $\ell_s$, the scales associated with the large $r$ behavior of $(u\cdot\chi)$ and $(s\cdot\chi)$ respectively, the scale of the bare cosmological constant $\ell$, and the scale of the effective cosmological constant, $\rt{|3/\La|}$, which determines the asymptotic behavior of the spacetime.

Only one of these scales is independent, owing to~\eq{c14=0:scales}, and we choose $\ell_s$ to be the independent parameter for the following reasons: $\ell_s$ can be understood in two ways. Most importantly, all $c_{14} = 0$ solutions have constant mean curvature hypersurfaces according to~\eq{sss:misc-expr} and~\eq{sss:central}, given by
	\beql{c14=0:K}
	K = -\fr{3}{\ell_s}~.
	\eeq
Therefore, $\ell_s$ determines the scale of the mean curvature of the constant {\ae}ther hypersurfaces. An additional physical way to interpret $\ell_s$ for the globally anti-de Sitter/Minkowski solutions, is that it is the {\ae}ther `mis-alignment' parameter. In the $c_{14} = 0$ global anti-de Sitter/Minkowski solutions the {\ae}ther can be aligned or mis-aligned with the global time-like Killing vector $\chi^a$. When the {\ae}ther is aligned with the Killing vector, $(s\cdot\chi) = 0$; this corresponds to the limit $\ell_s \to \infty$. Therefore, for all finite non-zero values of $\ell_s$, the {\ae}ther misalignment is appropriately parametrized by $\ell_s$. This misalignment changes the effective cosmological constant $\La$ via $\ref{c14=0:scales}$. Any anti-de Sitter solution with a finite $\ell_s$ is therefore characterized by a free parameter $\ell_s$, and we will express $\ell_u$ in terms of $\ell_s$ and various parameters of the Lagrangian using~\eq{c14=0:scales}, as shown in table~\ref{table:sss:global}. For the Minkowski case, $\ell_s = \ell_u$ always, and $\ell_s$ is a free parameter only if $c_{\CC} = 0$ and $c_l = 1$; for other choices of the couplings, the Minkowski solutions do not have any independent parameter.

In the case of global de Sitter solutions, we will still regard $\ell_s$ as the independent parameter, and express $\ell_u$ in terms of $\ell_s$ and the parameters of the Largangian as shown in table~\ref{table:sss:global}. However, in globally de Sitter solutions, the Killing vector must always have a non-zero $(s\cdot\chi)$ in the bulk of the spacetime, since it is spacelike outside the cosmological Killing horizon. Therefore, a deviation of $(u\cdot\chi)$ from $-1$ is a better measure of the {\ae}ther misalignment, and the parameter $\ell_u$ captures this best.
\subsection{Black hole solutions with maximally symmetric asymptotics}
All the globally maximally symmetric solutions presented in table~\ref{table:sss:global} above depend on at most one parameter. We now turn to the back hole solutions that are smoothly connected to each of them. As we shall see, these are always labeled by at least an additional mass parameter. For the special choice of $c_{123} = 0$, there is also an {\ae}ther charge necessary to describe the solution. Fundamentally this {\ae}ther charge arises because when $c_{123} = 0$, the action becomes insensitive to certain excitations of the {\ae}ther field. This leads to a gauge freedom which, in turn, has an associated conserved charge (c.f. related discussions in~\cite{Foster:disformal, JM:ae:intro}).

As already stressed, we will require that each black hole solution is smoothly connected to one of the globally maximally symmetric solution above. In other words, as the appropriate parameters/integration constants (e.g., the mass) are taken to zero, every solution reduces to a globally maximally symmetric solution. This requirement excludes solutions which cannot be thought of as excitations atop a maximally symmetric ground state. Solutions which do not satisfy our criteria/assumption, but which can nevertheless exist in principle, are potentially pathological in that it might not even be possible for example to define the mass of such solutions due to the necessity of a background subtraction~\cite{BY:defM, HH:defM}. It is also hard to see how such solutions could arise in a realistic situation, e.g., a collapse, where the initial state is presumably close to a maximally symmetric solution.
\subsubsection{Generic coupling case}
We will begin with the generic coupling case (the first column of table~\ref{table:sss:global}). In this case, the solutions can either be asymptotically de Sitter or asymptotically Minkowski. The corresponding asymptotic behavior of the functions $(u\cdot\chi)$, $(s\cdot\chi)$ and $f(r)$ are as in the left-most column of table~\ref{table:sss:global}. Furthermore, given their known asymptotic behavior one may solve for a series form of these functions in powers of $(1/r)$. In general, such series are asymptotic series, valid for `large $r$' in some suitable sense, and are not expected to converge globally. However, the analysis is sufficient to determine the maximum number of free parameters (integration constants) the solutions can depend on.
\paragraph{Asymptotically Minkowski:} For the asymptotically Minkowski case, such an analysis~\cite{BJS:aebh, EJ:aebh} reveals that the solutions can depend on at most two parameters, which we denote by $r_0$ and $c_{\ae}$. In particular, $r_0$ is the coefficient of the $\ord(1/r)$ term in the metric component $e(r)$. However, the two parameters appearing in the asymptotic analysis is actually an overcounting of the physical degrees of freedom. Numerical integration of the asymptotic expansions to reconstruct the bulk spacetime~\cite{BJS:aebh, EJ:aebh} reveals that the general two-parameter solutions are typically singular at the `spin-0 horizon'~\cite{EJ:aebh, BJS:aebh}, i.e., the trapped surface for the scalar excitations of the {\ae}ther with finite speed $s_0$~\eq{speeds}. Therefore, once the general solution is required to be regular everywhere outside the universal horizon, this extra constraint reduces the number of free parameters from two to one, which we choose to be $r_0$. The other parameter $c_{\ae}$ then becomes a function of the couplings $c_i$. Note that $c_{\ae}$ cannot depend on $r_0$, since $r_0$ is the {\it only} dimensionful quantity in the asymptotically Minkowski solutions while $c_{\ae}$ is dimensionless by definition. This observation will be centrally important in our derivation of the first law of mechanics for the above class of black holes in section~\ref{Smarr:1L}. We present the asymptotic form of the solution up to $\ord(r^{-5})$ in~\eq{generic-coupling:flat} below. 
	\begin{widetext}
	\beql{generic-coupling:flat}
	\begin{split}
	& (u\cdot\chi) = -1 + \fr{r_0}{2r} + \fr{r_0^2}{8r^2} + \fr{(6 + c_{14})r_0^3}{96r^3} + \lf[\fr{(5 + 2c_{14})}{128} -\fr{(1 - c_{13})c_{\ae}^2}{(2 - c_{14})}\rf]\fr{r_0^4}{r^4} + \ord(r^{-5})~, \\
	& (s\cdot\chi) = \fr{c_{\ae}r_0^2}{r^2} - \fr{c_{14}(1 - c_{13})c_{\ae}r_0^4}{40c_{123}r^4} + \ord(r^{-5})~, \\
	& f(r) = 1 + \fr{c_{14}r_0^2}{16r^2} + \fr{c_{14}r_0^3}{12r^3} + \fr{3c_{14}}{2}\lf[\fr{(16 + c_{14})}{256} -\fr{(1 - c_{13})c_{\ae}^2}{(2 - c_{14})}\rf]\fr{r_0^4}{r^4} + \ord(r^{-5})~, \\
	& e(r) = (u\cdot\chi)^2 - (s\cdot\chi)^2 = 1 - \fr{r_0}{r} - \fr{c_{14}r_0^3}{48r^3} - \lf[\fr{c_{14}}{48} + \fr{(2c_{13} - c_{14})c_{\ae}^2}{(2 - c_{14})}\rf]\fr{r_0^4}{r^4} + \ord(r^{-5})~.
	\end{split}
	\eeq
	\end{widetext}
\paragraph{Asymptotically de Sitter:} One can also perform a similar analysis for the asymptotically de Sitter solutions for the generic coupling case. The asymptotic forms of the solutions up to $\ord(r^{-5})$ are shown in~\eq{generic-coupling:dS} below (the scale of the effective cosmological constant is as in the global solution~\eq{c14!=0:dS:leff}: $\elef = \ell\rt{c_l}$)
	\begin{widetext}
	\beql{generic-coupling:dS}
	\begin{split}
	& (u\cdot\chi) = -1 + \fr{r_u}{r} - \fr{s_0^2r_u^2\elef^2}{6r^4} + \ord(r^{-5})~, \\
	& (s\cdot\chi) = \fr{r}{\elef} + \fr{(r_0 - 2r_u)\elef}{2r^2} + \fr{(2 - c_{14})r_u^2\elef}{4(1 - c_{13})r^3} + \ord(r^{-5})~, \\
	& f(r) = 1 - \fr{c_{14}s_0^2r_u^2\elef^2}{4r^4} + \ord(r^{-5})~, \\
	& e(r) = (u\cdot\chi)^2 - (s\cdot\chi)^2 = -\fr{r^2}{\elef^2} + 1 - \fr{r_0}{r} - \fr{(2c_{13} - c_{14})r_u^2}{2(1 - c_{13})r^2} - \fr{(2c_{13} - c_{14})s_0^2r_u^2\elef^2}{6(1 - c_{13})r^4} + \ord(r^{-5})~.
	\end{split}	
	\eeq
	\end{widetext}
From~\eq{generic-coupling:dS}, one may see that these solutions also depend on two parameters, which have been called $r_0$ and $r_u$ here. The parameter $r_0$, as before, is the $\ord(1/r)$ coefficient of $e(r)$ with similar interpretation. The other parameter, $r_u$, can either be thought as coming from the $\ord(1/r)$ term of $(u\cdot\chi)$ or the $\ord(1/r^2)$ term of $(s\cdot\chi)$. Also note that some of the higher order coefficients can be written in a compact manner in terms of the spin-0 mode speed $s_0$~\eq{speeds}. We expect that the solution will be generally singular at the spin-0 horizon, and demanding global regularity of the solution from the universal horizon and beyond will fix $r_u$ as a function of $r_0$ and $\ell$. However, to the best of our knowledge, a thorough analysis clarifying the spin-0 horizon regularity for the asymptotically de Sitter solution has not been carried out. Such an analysis must eventually occur if one wants to examine black hole thermodynamics in asymptotically de Sitter spacetimes, as the regularity condition changes the number of asymptotic charges/parameters.
\subsubsection{$c_{123}=0$}
Both the $c_{123}=0$ and $c_{14}=0$ cases require $f(r) = 1$\ft{For either of these choices of the couplings, the `cross-component' of the {\ae}ther stress tensor $\aePi_s$ vanish, which requires $\Ric_{u s}$ to vanish on shell. The only non-trivial way to the satisfy the latter is to require $f(r) =$ constant, which can be set to one by boundary condition.}, and one can then explicitly integrate the remaining equations and obtain analytic forms of the solutions.

For $c_{123} = 0$, the complete solution, encompassing both the asymptotically flat as well as de Sitter cases, is as follows~\cite{JB:thesis}
	\beql{c123=0}
	\begin{split}
	& (u\cdot\chi) = -1 + \fr{\ruh}{r}~, \\
	& (s\cdot\chi) = \rt{\fr{c_{\CC}r^2}{\elef^2} + \fr{r_0 - 2\ruh}{r} + \fr{(2 - c_{14})\ruh^2}{2(1 - c_{13})r^2}}~, \\
	& f(r) = 1~, \\
	& e(r) = 1 - \fr{c_{\CC}r^2}{\elef^2} - \fr{r_0}{r} + \fr{(c_{14} - 2c_{13})}{2(1 - c_{13})}\fr{\ruh^2}{r^2}~,
	\end{split}
	\eeq
where $c_{\CC} = 0, 1$ for the asymptotically flat and de Sitter cases respectively and $\elef = \ell\rt{1 - c_{13}}$ for the asymptotically de Sitter case~\eq{c14!=0:dS:leff}. Note that unlike the generic coupling Minkowski and de Sitter cases, here the solution depends on {\it two} parameters/integration constants $r_0$ and $\ruh$. The latter is the radial location of the universal horizon in these solutions, as can be inferred from the expression for $(u\cdot\chi)$. In particular, the explicit choice $\ruh > 0$ has been made to ensure that a universal horizon exists and cloaks the curvature singularity at $r = 0$~\cite{JB:thesis}. For asymptotically flat solutions, one must furthermore demand $r_0 \geqq 2\ruh$ for a non-singular behaviour of $(s\cdot\chi)$ as $r \to \infty$. For asymptotically de Sitter solutions, this is not necessary, but there does exist a lower bound on $r_0 - 2\ruh$; we will refrain from showing the exact relations here due to their involved nature and minimal relevance for the rest of the discussion. To avoid a naked singularity one must also demand that $r_0 \to 0$ as $\ruh \to 0$. Finally, one may verify that in the said limit, the above solution smoothly reduces to the global de Sitter solution for $c_{123} = 0$~\eq{c14!=0:global:dS} when $c_{\CC} = 1$, while to the global Minkowski solution for $c_{123} = 0$~\eq{c14!=0:global:flat} when $c_{\CC} = 0$. Note that a special case of this solution, corresponding to no bare cosmological constant ($c_{\CC} = 0$) and the special choice of parameters $r_0 = 2\ruh$ already appeared in~\cite{BBM:mechuhor}.

The appearance of the extra charge, $\ruh$, compared to the generic coupling case can be understood by thinking about the broader context of Einstein-{\ae}ther/Ho\v{r}ava-Lifshitz lagrangians and their associated solution spaces. As pointed out in~\cite{Foster:disformal}, solution spaces of Einstein-{\ae}ther theory are invariant under field redefinitions consisting of a disformal transformation of the metric $\met_{a b} \to \Om^2\met_{a b} + (\Om^2 - \Om_u^2)u_a u_b$ along with a scaling of the {\ae}ther $u^a \to \Om_u^{-1}u^a$ that maintains the unit norm constraint~\eq{ae:norm}, where $\Om$ and $\Om_u$ are constants. In Ho\v{r}ava-Lifshitz gravity and the twist-free sector of Einstein-{\ae}ther gravity the twist-free condition on the {\ae}ther is maintained by these transformations. More importantly, the form of the action remains invariant under such transformations if some of the coefficients transform accordingly. In particular, $(1 - c_{13})$ and $(1 + c_2)$ are multiplied by a factor of $\Om_u^2\Om^{-2}$, $c_{14}$ stays invariant, the scale of the bare cosmological constant picks up a factor of $\Om$ and the normalization constant $\Gae$ gets a factor of $\Om_u\Om$~\cite{Foster:disformal}. Note that these couplings are actually the ones that appear in the Ho\v{r}ava-Lifshitz action~\eq{eq:dictionary}.

As a result, the forms of the equations of motion are unchanged as well, and the number of asymptotic charges that describe the solutions does not change under a non-singular field redefinition of the above form either. Now, since $c_{123} = 0$ is preserved under these transformations, an appropriate choice of $\Om$ and $\Om_u$ that allows one to set $c_{13} = 0$ makes the $c_{123} = 0$ solution space equivalent to one where $c_{13} = c_2 = 0$. Using~\eq{Du:decomp:HSO}, one may then note the algebraic relation $F_{a b}F^{a b} = -2a^2$, where $F_{a b} = 2\Dl_{[a} u_{b]}$, and express the action~\eq{ae:action} with~\eq{ae:lag:HSO} as (see also~\cite{Foster:disformal})
	\beqn{
	\ac = \fr{1}{16\pi\Gae}\intdx{4}\rt{-\met}\lf[-\fr{6c_{\CC}}{\ell^2} + \Rie - \fr{c_{14}}{2}F_{a b}F^{a b}\rf],
	}
upto the constraint term enforcing the unit norm~\eq{ae:norm}. We recognize this as a (partially) gauge fixed Einstein-Maxwell theory, where the gauge is such that the vector potential is unit normal. The gauge fixing does not completely specify the corresponding $U(1)$ gauge, as any transformation of $u^a$ that preserves the constraint~\eq{ae:norm} is allowed~\cite{JM:ae:intro}. Hence there is an extra residual gauge freedom and an extra charge as long as the {\ae}ther is hypersurface orthogonal. The `Coulombic' fall-off of the acceleration in these solutions
	\beqn{
	(a\cdot s) = \fr{\ruh}{r^2}~,
	}
allows us to recognize the acceleration vector as playing the role of `electric field components' of $F_{a b}$ and confirms $\ruh$ as the corresponding conserved charge.
\subsubsection{$c_{14}=0$}
Finally, let us turn to the general solutions for the remaining case of $c_{14} = 0$. These solutions can also be presented in a closed analytic form, with the most general solution given as follows
	\begin{widetext}
	\beql{c14=0}
	\begin{split}
	& (u\cdot\chi) = -\fr{r}{\ell_u}\lf(1 -\fr{\ruh}{r}\rf)\rt{1 + \fr{2\ruh}{r} + \fr{3\ruh^2 + \ell_u^2}{r^2} + \fr{2\ruh(3\ruh^2 + \ell_u^2)}{3r^3} + \fr{\ruh^2(3\ruh^2 + \ell_u^2)}{3r^4}}~, \\
	& (s\cdot\chi) = \fr{r}{\ell_s} + \fr{\ruh^2}{r^2\rt{3(1 - c_{13})}}\rt{1 + \fr{3\ruh^2}{\ell_u^2}}~, \\
	& e(r) = -\fr{\La r^2}{3} + 1 - \fr{r_0}{r} - \fr{c_{13}}{3(1 - c_{13})}\lf(1 + \fr{3\ruh^2}{\ell_u^2}\rf)\fr{\ruh^4}{r^4}~, \\
	& f(r) = 1~, \\
	& r_0 = \fr{4\ruh}{3} + \fr{2\ruh^3}{\ell_u^2} + \fr{2\ruh^2}{\ell_s\rt{3(1 - c_{13})}}\rt{1 + \fr{3\ruh^2}{\ell_u^2}}~.
	\end{split}
	\eeq
	\end{widetext}
Apart from the dependence on $\ell_s$, which already appears in the global solution~\eq{c14=0:global}, the general solution~\eq{c14=0} depends on one more parameter $\ruh$ -- the location of the universal horizon in these solutions. As before, the choice $\ruh > 0$ has been made to hide a curvature singularity at $r = 0$ behind the universal horizon. Indeed, one may take the $\ruh \to 0$ limit of the above to recover~\eq{c14=0:global}. Using~\eq{sss:misc-expr}, one may further confirm that the mean curvature $K$ of the preferred foliations for the above solution is constant and given by~\eq{c14=0:K}. The effective cosmological constant $\La$ is as in the global solution~\eq{c14=0:scales}, and the asymptotic behavior is determined by the relatives sizes of $\ell_u$ and $\ell_s$~\eq{c14=0:asymp}. As with the other solutions discussed before, the $\ord(1/r)$ coefficient of $e(r)$, which depends on $\ruh$ as shown above, is related to the mass of solutions. However, since $r_0$ and $\ruh$ are directly related there is no additional charge associated with the black hole. Above a given global background characterized by a fixed $\ell_s$, there is only a one parameter family of black holes, unlike the $c_{123}=0$ case, where there is a two parameter family of black holes. The special class of asymptotically flat solutions obtained by taking the simultaneous limits $\ell_u,\,\ell_s \to 0$ has been reported previously in~\cite{BBM:mechuhor}.
\section{Smarr and first law formulae}\label{Smarr:1L}
In general relativity, Smarr formulae~\cite{Smarr} relate the asymptotic charges of a black hole to quantities defined on its Killing horizon. Mathematically, such a relation exists due to general covariance that general relativity enjoys~\cite{BCH, Wald:NQ}. Furthermore, upon considering variations between `nearby solutions' (i.e., solutions differing at the first order in their parameters), the corresponding variation of the Smarr formula gives the first law of black hole mechanics~\cite{BCH}. General covariance does not, however, automatically yield a Smarr formula that has a manifestly thermodynamic analog. Indeed, applying the Noether approach for a first law to a Killing horizon in Einstein-{\ae}ther theory for asymptotically flat black holes yields a first law without any natural thermodynamic analog~\cite{Foster:NQ}. As has been shown in~\cite{BBM:mechuhor, JB:thesis}, however, evaluating the first law and Smarr on the universal horizon can give a first law with a natural thermodynamic interpretation. The goal of this section is to first provide a simple derivation of a Smarr formula for general static and spherically symmetric {\ae}ther black holes, and subsequently obtain the corresponding first law of black hole mechanics with respect to the universal horizon.
\subsection{General approach}
One can quickly derive a Smarr formula if one can construct a divergence free two form in the bulk, which upon integration will give rise to a Smarr formula relating quantities on the universal horizon and asymptotic infinity~\cite{BCH, Wald:NQ}. Following~\cite{BBM:mechuhor, JB:thesis} one can construct such a two-form $\CMcal{F}_{a b}$ as follows
	\beql{Smarr:local}
	\CMcal{F}_{a b} = \qsmarr(r)\veii_{a b}~, \qquad \Dl_b\CMcal{F}^{a b} = 0~.
	\eeq
Here $\veii_{a b} \equiv 2u_{[a} s_{b]}$, $\qsmarr(r)$ is given by
	\beql{def:qsmarr}
	\begin{split}
	\qsmarr(r) & = q_{\CC}(r) - \lf(1 - \fr{c_{14}}{2}\rf)(a\cdot s)(u\cdot\chi) \\
	& \quad + (1 - c_{13})K_{s s}(s\cdot\chi) + \fr{c_{123}}{2}K(s\cdot\chi)~,
	\end{split}
	\eeq
and $q_{\CC}(r)$ is implicitly defined through
	\beql{qcc:ODE}
	\lf[r^2q_{\CC}(r)\rf]' = \fr{3c_{\CC}}{\ell^2}r^2f(r)~.
	\eeq
Note that the proportionality between $\CMcal{F}_{a b}$ and $\veii_{a b}$ is purely a consequence of spherical symmetry.

Given any explicit solution, one may integrate~\eq{qcc:ODE} to solve for $q_{\CC}(r)$. In particular, when $f(r) = 1$ one readily has
	\beql{qcc:f=1}
	q_{\CC}(r) = \fr{c_{\CC}r}{\ell^2}.
	\eeq
This form of $q_{\CC}(r)$ is therefore relevant for all the globally maximally symmetric solutions listed in table~\ref{table:sss:global}. Strictly speaking, for $f(r) = 1$, $q_{\CC}(r)$ is given as in~\eq{qcc:f=1} only up to some constant$\times r^{-2}$, since any solution of~\eq{qcc:ODE} determines $r^2q_{\CC}(r)$ up to some constant of integration. Setting this constant to zero is physically equivalent to imposing the requirement that the mass of a globally maximally symmetric solution (`reference background') is zero, i.e., all mass quantities are relative to the appropriate `background subtraction'~\cite{BY:defM, HH:defM}. We also emphasize that this arbitrary constant of integration does not affect the linearly divergent nature of $q_{\CC}(r)$ for large $r$ (i.e., if $c_{\CC} \neq 0$ in~\eq{qcc:f=1}). Rather, as one may verify explicitly, the combination of $(u\cdot\chi)$, $(s\cdot\chi)$, $(a\cdot s)$, $K_{ss}$ and $K$ as appearing in $\qsmarr(r)$~\eq{def:qsmarr} diverges in precisely the correct way as to cancel the corresponding divergence in $q_{\CC}(r)$.

An additional subtlety arises for the $c_{14}=0$ solutions. As has already been noted, for $c_{14} = 0$~\eq{c14=0} there exist globally maximally symmetric solutions characterized by a non-zero bare and/or effective cosmological constant, generated by a combination of the bare cosmological constant term in the action and an {\ae}ther profile where the {\ae}ther is misaligned with respect to the time-like Killing vector $\chi^a$. We also have black hole solutions~\eq{c14=0} that are smoothly connected to these global solutions. In particuar, such solutions can be generated even without a cosmological constant term in the action. While $q_{\CC}(r) = 0$ for such solutions the corresponding divergences in the {\ae}ther profile for large $r$ still cancel out appropriately in $\qsmarr(r)$. In fact, $\qsmarr(r)$ as defined above (with the integration constant controlling $r^{-2}$ behavior in $q_{\CC}(r)$ set to zero) vanishes identically for all globally maximally symmetric backgrounds, even those with cosmological constants generated solely by a misaligned {\ae}ther. Hence all our Smarr formula are already background subtracted -- the infinite energy of the corresponding globally maximally symmetric solutions, specified by a non-zero cosmological constant (bare and/or effective), does not appear.

As a consequence of the inherent background subtraction in our approach, we must consider any parameters such as $\ell_s$ that control both the black hole solutions and the globally maximally symmetric solutions as fixed when deriving first laws from the Smarr formulae. This allows us to generate the first law for black holes on top of a given background solution. In particular, we will not vary $\ell_s$. It may be interesting in future work to consider variations of $\ell_s$, as there is a program in general relativity to examine the effect of a variable cosmological constant on the first law~\cite{Traschen} which is similar to variations of $\La$, the effective cosmological constant in our solutions.

For the corresponding black hole solutions, note that a specification of $q_{\CC}(r)$ for each and every case has to be made consistently with the corresponding global solution, since we want to consistently subtract the background energy~\cite{BY:defM, HH:defM}. This is especially relevant for situations with a bare cosmological constant, and/or with a non-trivial {\ae}ther profile where the {\ae}ther is misaligned with respect to the Killing vector $\chi^a$ asymptotically. With this in mind, $q_{\CC}(r) = 0$ for the asymptotically flat black hole solution for generic values of the couplings~\eq{generic-coupling:flat}, while for the exact black hole solutions for $c_{123} = 0$~\eq{c123=0} or $c_{14} = 0$~\eq{c14=0}, $q_{\CC}(r)$ is given as in~\eq{qcc:f=1}. On the other hand, for the asymptotically de Sitter black hole solutions for generic coupling~\eq{generic-coupling:dS}, one could only present a series expansion of $q_{\CC}(r)$ in powers of $(1/r)$ with the final result as follows
	\beql{qcc:dS}
	q_{\CC}(r) = \fr{r}{\ell^2} + \fr{3c_{123}(2 - c_{14})r_u^2}{8(1 - c_{13})r^3} + \ord(r^{-5})~.
	\eeq
Note that the leading order (divergent) behaviour is exactly as in~\eq{qcc:f=1} as expected. 

The leftover finite piece in $\qsmarr(r)$ for a black hole solution can now be easily determined by noting that according to~\eq{Smarr:local}, $\CMcal{F}_{a b}$ behaves like the `radial electric field due to a point charge at the origin'. Consequently,
	\beql{qsmarr:finite}
	\qsmarr(r) = \fr{\ell_{\ae}(\{c_i\}, \ruh, \cdots)}{2r^2}~,
	\eeq
where $\ell_{\ae}(\{c_i\}, \ruh, \cdots)$ is a constant with the dimension of length, which generically depends on the couplings $\{c_i\} \equiv \{c_2, c_{13}, c_{14}\}$, the parameter $\ruh$ which labels every solution, the scale $\ell$ of the cosmological constant if present, and possibly other parameters (e.g., $r_0$, $\ell_u$ and/or $\ell_s$ as and when applicable). Note that $\ell_{\ae} = 0$ whenever $\ruh = 0$ by our discussion above. For the actual black hole solutions, we will present the forms of $\ell_{\ae}(\{c_i\}, \ruh, \cdots)$ on a case-by-case basis below.

Even with $\ell_{\ae}(\{c_i\}, \ruh, \cdots)$ unspecified, however, we can compute the flux of $\CMcal{F}_{a b}$ through any two-sphere $\Sph_r$ at a radius $r$, simply by integrating~\eq{Smarr:local}. In fact, due to the `electric-field' like behaviour of $\CMcal{F}_{a b}$ (equivalently, since $\qsmarr(r) \sim r^{-2}$), the flux is independent of $r$ and hence the flux through `the boundary at infinity' equals the flux through the universal horizon. Therefore, in terms of the {\it total mass} $M_{\ae}$ of any black hole solution, which we {\it define} as
	\beql{def:Mae}
	M_{\ae} = \fr{\ell_{\ae}(\{c_i\}, \ruh, \cdots)}{2\Gae}~,
	\eeq
the statement of equality of the fluxes through the asymptotic boundary and the universal horizon is
	\beql{Smarr:global}
	M_{\ae} = \fr{\quh A_{\uhor}}{4\pi\Gae}~,
	\eeq
where $\quh \equiv \qsmarr(\ruh)$ and $A_{\uhor} \equiv 4\pi\ruh^2$ is the area of the universal horizon. The above relation is the sought-after Smarr formula, valid for {\it all} black hole solutions presented above. Upon recalling~\eq{def:qsmarr}, $\quh$ can be written in terms of more familiar quantities as
	\beql{def:q_uh}
	\quh = q_{\CC}(\ruh) + (1 - c_{13})\kauh + \fr{c_{123}}{2}K_{\uhor}\Xuh~, 
	\eeq
where $\kauh$ and $K_{\uhor}$ are the surface gravity~\eq{def:ka} and the trace of the extrinsic curvature $K$ evaluated on the universal horizon, respectively, and $\Xuh$ is the magnitude of the Killing vector on the universal horizon (recall, $(s\cdot\chi)_{\uhor} = \Xuh$~\eq{u:EF}).

Varying the Smarr formula~\eq{Smarr:global} yields a first law of black hole mechanics. Physically, such a variation takes us from one regular and static solution (labeled by a set of parameters) to a distinct nearby regular and static solution, such that the parameters have changed `infinitesimally'. In practice, this means that we need to consider the first order variation of both sides of~\eq{Smarr:global} generated by a variation of the underlying parameters. Such a variation can be computed directly if the analytical form of the solution under consideration is known explicitly (more precisely, we only need to know the explicit dependence of both sides of~\eq{Smarr:global} on the parameters). This is true for the exact solutions~\eq{c123=0} and~\eq{c14=0}. Furthermore, for the asymptotically flat solution for generic values of the coupling~\eq{generic-coupling:flat}, the scaling argument presented in~\cite{BBM:mechuhor} (to be reviewed below) is sufficient to obtain a first law for this case. Unfortunately, for the asymptotically de Sitter solution for generic couplings~\eq{generic-coupling:dS} (and only for this case) none of the above strategies can work, due to a lack of complete knowledge of the solution and the presence of a dimensionful constant in the solution -- the scale of the bare cosmological constant.
\subsection{Generic coupling case}
\subsubsection{Asymptotically flat solutions}
Let us begin with the asymptotically flat black hole solutions for generic coupling case~\eq{generic-coupling:flat}. Here, even though we do not have exact solutions, the boundary behavior of the metric and the {\ae}ther profile leads to
	\beql{Mae:gen:flat}
	M_{\ae} = \lf(1 - \fr{c_{14}}{2}\rf)\fr{r_0}{2 G_{\ae}}~.
	\eeq
From our knowledge of $r_0$ as the coefficient of the $1/r$ term in the metric expansion, we recognize the term not proportional to $c_{14}$ as simply the ADM mass of the solutions. The extra term is the {\ae}ther contribution to the total energy -- the combined expression matches what has been found previously for the total energy~\cite{Eling:Mae, Foster:NQ, GJ:+veEae}.

Due to our lack of knowledge of the analytical form of the solution, we do not know the explicit functional relationship between $M_{\ae}$ (or $r_0$) and $\ruh$. However, since the solution depends on a single dimensionful parameter, we must have $M_{\ae} \propto \ruh$ with the constant of proportionality a function of $c_2$, $c_{13}$ and $c_{14}$ {\it but of nothing else}. We also can argue on similar grounds that $\quh \propto \ruh^{-1}$. Finally, since $A_{\uhor} \propto \ruh^2$, variations of $\quh$ and $A_{\uhor}$, due to that of $\ruh$, are related by
	\beqn{
	\de\quh A_{\uhor} = -\fr{\quh\,\de A_{\uhor}}{2}~.
	}
So, upon considering a variation of~\eq{Smarr:global} for the present solutions, we get
	\beql{1L:gen-flat}
	\de M_{\ae} = \fr{\quh\,\de A_{\uhor}}{8\pi\Gae}~.
	\eeq
This is, therefore, the first law for the generic asymptotically flat case, originally presented as above in~\cite{BBM:mechuhor}. We emphasize the crucial role played by the single dimensionful parameter dependence of the solutions, due to which, even without a complete knowledge of the solutions, we could obtain a first law.
\subsubsection{Asymptotically de Sitter}
We are not so fortunate in the the case of asymptotic de Sitter solutions~\eq{generic-coupling:dS} as there is now a second scale in the problem -- that of the bare cosmological constant. Hence we will not be able to derive a general first law for this class. One can still quickly compute the total mass from~\eq{def:Mae},
	\beql{Mae:gen:dS}
	M_{\ae} = \lf[\lf(1 - \fr{c_{14}}{2}\rf)r_u + \fr{(1 - c_{13} - \fr{3}{2}c_{123})(r_0 - 2r_u)}{2}\rf]\fr{1}{\Gae}.
	\eeq
As discussed previously, $r_u$ is in principle related to $r_0$ by requiring regularity of the spin-0 horizon in the bulk. However a direct linear relationship cannot be argued for as the presence of the effective cosmological constant length scale $\elef = \ell\rt{c_l}$ allows for arbitrary functional forms of $r_0/\elef$. Similarly, one cannot argue that $M_{\ae} \propto \ruh$ and hence one cannot derive a first law relating a variation at the universal horizon to a variation of the total mass. The situation improves dramatically when $c_{123}=0$ or $c_{14}=0$ as we have complete analytic solutions.
\subsection{$c_{123}=0$}
Let us next look at the black hole solutions for $c_{123} = 0$~\eq{c123=0}. We will explicitly consider the cases with a universal horizon (i.e., $\ruh \neq 0$). From the knowledge of the solutions, a direct computation, using~\eq{def:Mae}, yields $M_{\ae}$
	\beql{Mae:c123=0}
	\begin{split}
	M_{\ae} & = \lf[\lf(\fr{1 - c_{13}}{2}\rf)r_0 + \lf(c_{13} - \fr{c_{14}}{2}\rf)\ruh\rf]\fr{1}{\Gae}~, \\
	& = \lf(1 - \fr{c_{14}}{2}\rf)\fr{r_0}{2\Gae} + \fr{(c_{14} - 2c_{13})(r_0 - 2\ruh)}{4\Gae}~.
	\end{split}
	\eeq
We caution the reader about the following difference between these solutions and black hole solutions in general relativity. Na\"ively in the metric $\ruh$ plays the role of the electric `charge' while $r_0$ plays the role of the (ADM) mass. However, as seen above, there is a contribution to the mass from the {\ae}ther which involves $\ruh$ as well. Furthermore, the location of the universal horizon depends solely on $\ruh$. This is very different than the general relativistic case, where the mass depends only on $r_0$ and the radius of the Killing horizon depends on both the mass and the charge. Varying the mass formula~\eq{Mae:c123=0} with respect to $\ruh$ and $r_0$ yields
	\beql{1L:c123=0}
	\de M_{\ae} = \fr{{\bar{q}_{\uhor}}\,\de A_{\uhor}}{8\pi\Gae} + \fr{(1 - c_{13})\de r_0}{2\Gae}~,
	\eeq
where 
	\beql{eq:quhc123eq0}
	\bar{q}_{\uhor} = \fr{1}{\ruh}\lf(c_{13} - \fr{c_{14}}{2}\rf)~.
	\eeq
Note that one can make contact with the form of the Smarr formula in (\ref{Smarr:global}) by relating $\quh$ and $\bar{q}_{\uhor}$ via
	\beqn{
	\quh = \bar{q}_{\uhor} + \fr{(1 - c_{13})r_0}{2\ruh^2}~.
	}
This gives us the first law of black hole mechanics for these solutions. A thermodynamic interpretation of this first law, which is evidently quite different from general relativity due to the presence of the $\de r_0$ term, we leave to future work.
\subsection{$c_{14}=0$}
The final class of black hole exact solutions were given by the generally two-parameter family of solutions~\eq{c14=0}. Once again, we will consider the cases with $\ruh \neq 0$. A direct computation using~\eq{def:Mae} yields
	\beql{Mae:c14=0}
	\begin{split}
	M_{\ae} = \Biggl[& \fr{2\ruh}{3} + \fr{\ruh^3}{\ell_u^2} \\
	& + \fr{[2(1 - c_{13}) - c_l]\ruh^2}{\ell_s\rt{3(1 - c_{13})}}\rt{1 + \fr{3\ruh^2}{\ell_u^2}}\Biggr]\fr{1}{\Gae}~,
	\end{split}
	\eeq
which is a function of $\ruh$ and $\ell_s$, the latter both explicitly as well as through $\ell_u$.

When deriving a first law from the mass formula~\eq{Mae:c14=0}, we recall that, as discussed previously, we should only take variations with respect to $\ruh$. The parameter $\ell_s$ (and therefore $\ell_u$ as well) also describes the globally maximally symmetric background solution and hence should not be varied. Varying the mass formula~\eq{Mae:c14=0} with respect to $\ruh$ yields
	\beql{1L:c14=0}
	\de M_{\ae} = \fr{\bar{q}_{\uhor}\de A_{\uhor}}{8\pi\Gae}~,
	\eeq
where
	\beqn{
	\begin{split}
	\bar{q}_{\uhor} = & \fr{2}{3\ruh} + \fr{3\ruh}{\ell_u^2} \\
	& + \fr{[2 - (5c_{13} + 3c_2)]}{\ell_u\rt{3(1 - c_{13})}}\lf[1 + \fr{9\ruh^2}{2\ell_u^2}\rf]\lf[1 + \fr{3\ruh^2}{\ell_u^2}\rf]^{-\fr{1}{2}}.
	\end{split}
	}
As a consistency check, note that the limit with $\ell,\,\ell_s \to \infty$ should correspond to the asymptotically flat case with the {\ae}ther aligned with the time-like Killing vector at infinity. In this limit $\ell_u$ also goes to infinity, which reduces the first law to 
	\beqn{
	\de M_{\ae} = \fr{2}{3\ruh}\fr{\de A_{\uhor}}{8\pi\Gae}~,
	}
which matches the $c_{14}=0$ first law previously found via the conserved two-form approach~\cite{BBM:mechuhor} and the Noether approach~\cite{Mohd:2013zca}. 
	
While the thermodynamic role for the last term in (\ref{1L:c14=0}) is not immediately obvious, the first two terms in the first law may have a thermodynamic interpretation that meshes nicely with what happens in general relativity. The first term is equivalent to the usual $\kappa \de A$ term in general relativity, while the second term can be interpreted as a work term, $P \de V$, with the effective pressure $P$ controlled by $\ell_u$. Such pressure terms have been considered for anti-de Sitter black holes in general relativity~\cite{Traschen}, and $\ell_u$ is related to the effective cosmological constant so such an interpretation seems plausible. However, due to the presence of the last term, the full thermodynamic interpretation is not naively clear and requires further investigation. 
\section{Conclusion}
The existence of universal horizons and the corresponding Smarr formulae, first laws, and apparent thermal radiation in special cases~\cite{BBM:thermo}, suggest that black hole thermodynamics has a significant extension into theories that have different causal structures than general relativity. As noted by Schwarzschild, in any situation with new and poorly understood physics ``it is always pleasant to have exact solutions in simple form at your disposal." In this paper we have presented a number of exact solutions for both globally maximally symmetric spacetimes as well as static and spherically symmetric spacetimes with universal horizons and maximally symmetric asymptotics, and given asymptotic expansions for similar solutions where a closed analytic form cannot be found. We have further provided a complete classification of physically relevant static and spherically symmetric black hole solutions by both the asymptotic form and the coefficients in the Lagrangian. In addition, we proved that the solution spaces for static and spherically symmetric spacetimes with universal horizons in Ho\v{r}ava-Lifshitz theory and Einstein-{\ae}ther theory are isomorphic, making it clear that with these symmetries one can work in either theory (unlike with the rotating solution spaces, which are quite different~\cite{BS:aebh:R1,BS:aebh:R2}). We should also point out that none of our solutions are expected to admit `maximal extensions' as in general relativity, due to singular behaviour of the {\ae}ther on various null hypersurfaces, as noted in~\cite{Sotiriou:2014gna}.

Beyond the solutions, we also generated Smarr formulae and the first law for each universal horizon solution. Of interest is that anti-de Sitter spacetimes are only possible when the $c_{14}$ coefficient in the Einstein-{\ae}ther lagrangian vanishes. This is an issue for any eventual thermodynamic interpretation. Asymptotically anti-de Sitter spaces are the best understood from a holographic perspective, yet the $c_{14} = 0$ subset of Einstein-{\ae}ther Lagrangians has been argued to be poorly behaved in terms of generic solutions~\cite{Henneaux:2009zb}. In general relativity, we are used to black hole solutions being the end points of very generic initial conditions.  However, since our solutions are very specific cases, they may exist and be well behaved, but not achievable as end states of generic collapse as in general relativity. The work of Afshordi et. al.~\cite{Saravani:2013kva} shows that universal horizons can be end points of spherical collapse, but this is still a very special case. Such dynamical issues are not necessarily relevant for counting black states or determining thermodynamic relations, but they must eventually be understood in a complete thermodynamical interpretation for universal horizons, as the end result of collapse being a black hole meshes neatly with the second law.

The asymptotically anti-de Sitter universal horizon physics also shows differences from the general relativistic case in that specifying global anti-de Sitter geometry does not uniquely specify the appropriate background solution for a universal horizon. One must also specify the alignment of the {\ae}ther vector which influences the form of the first law for universal horizons.

There are many issues that remain. A thermodynamic interpretation of the first law must still be proven, and it is unclear how terms that do not appear in the general relativistic case should be interpreted. In the de Sitter case, the na\"ive two parameter asymptotic expansion is expected to reduce to a one parameter family once the spin-0 horizon in the bulk is required to be regular. However, currently there is no known analytic approach to determine the relation between the parameters in the solution, which makes a first law impossible to derive in these cases. Finally, global Lifshitz spacetimes have been shown to be solutions of Ho\v{r}ava-Lifshitz and Einstein-{\ae}ther theory~\cite{Griffin:2012qx}. Whether such solutions also admit universal horizons is an important open question as asymptotically Lifshitz spacetimes are extensively used to study duals to condensed matter systems. Our hope is that the solutions and first laws presented in the work serve as useful tools for investigating the rich and deep physics behind universal horizons, just as the Schwarzschild solution has been a terribly important tool in investigating black hole thermodynamics in general relativity.
\acknowledgements{The authors thank Thomas Sotiriou, Sayandeb Basu, Ted Jacobson, Mattia Colombo, and Per Berglund for useful conversations and feedback. DM thanks the University of New Hampshire for research support. JB thanks the University of New Hampshire, USA and SISSA, Trieste, Italy for hospitality where part of the research was performed. JB also acknowledges financial support from the European Research Council under the European Union's Seventh Framework Programme (FP7/2007-2013) / ERC Grant Agreement n. 306425 “Challenging General Relativity”.}

\end{document}